\documentclass[fleqn,usenatbib]{mnras}

\usepackage[T1]{fontenc}
\usepackage{ae,aecompl}

\usepackage{graphicx}	
\usepackage{amsmath}	
\usepackage{amssymb}	
\usepackage{bigints}

\usepackage{xcolor}     
\usepackage[normalem]{ulem}
\usepackage{hyperref}
\usepackage{comment}

\usepackage{newtxtext,newtxmath}

\newlength{\VSpaceBeforeTabBib}
\newlength{\VSpaceBeforeTabFoot}
\newcommand*\tablefootname{Notes}
\newcommand*\tablefootfont{\small}
\newcommand*\tablefootnamefont{\small\bfseries}
\newcommand\tablefoot[1]{\VSpaceBeforeTabBib=1ex%
  \par\vspace{\VSpaceBeforeTabFoot}
  \noindent
  \begin{minipage}{\linewidth}
    {\tablefootnamefont\tablefootname.}~%
    \tablefootfont
    \ignorespaces
    #1%
  \end{minipage}%
}
\newcommand*\tablefootmark[1]{%
  \unskip
  \hbox{\textsuperscript{\normalfont\itshape\ignorespaces#1}}%
  \,%
  \ignorespaces
}
\newcommand\tablefoottext[2]{%
  \hbox{\textsuperscript{\normalfont({\itshape\ignorespaces#1})}}%
  ~%
  \ignorespaces
  #2\ \ignorespaces%
}

\title[Envelope-Disk Transition of Dust Polarization]
{The Transition of Polarized Dust Thermal Emission from the Protostellar Envelope to the Disk Scale}

\author[K. H. Lam et al.]{
Ka Ho Lam,$^{1}$\thanks{E-mail: kl4sf@virginia.edu}
Che-Yu Chen,$^{1,2}$
Zhi-Yun Li,$^{1}$
Haifeng Yang,$^{3}$
Erin G. Cox,$^{4}$
\newauthor
Leslie W. Looney,$^{5}$
and Ian Stephens$^{6,7}$
\\
$^{1}$Department of Astronomy, University of Virginia, 530 McCormick Rd., Charlottesville 22904, Virginia, USA\\
$^{2}$Lawrence Livermore National Laboratory, 7000 East Ave, Livermore, CA 94550, USA\\
$^{3}$Institute for Advanced Study, Tsinghua University, Beijing 100084, People's Republic of China\\
$^{4}$Center for Interdisciplinary Exploration and Research in Astrophysics, Northwestern University, 1800 Sherman Ave, Evanston, IL 60201, USA\\
$^{5}$Department of Astronomy, University of Illinois, 1002 West Green Street, Urbana, Illinois 61801, USA\\
$^{6}$Harvard-Smithsonian Center for Astrophysics, 60 Garden Street, Cambridge, MA 02138, USA\\
$^{7}$Department of Earth, Environment and Physics, Worcester State University, Worcester, MA 01602, USA\\
}

\date{Accepted XXX. Received YYY; in original form ZZZ}

\pubyear{\the\year}

\begin{document}
\label{firstpage}
\pagerange{\pageref{firstpage}--\pageref{lastpage}}
\maketitle

\begin{abstract}

Polarized dust continuum emission has been observed with ALMA in an increasing number of deeply embedded protostellar systems. It generally shows a sharp transition going from the protostellar envelope to the disk scale, with the polarization fraction typically dropping from ${\sim} 5\%$ to ${\sim} 1\%$ and the inferred magnetic field orientations becoming more aligned with the major axis of the system. We quantitatively investigate these observational trends using a sample of protostars in the Perseus molecular cloud and compare these features with a non-ideal MHD disk formation simulation. We find that the gas density increases faster than the magnetic field strength in the transition from the envelope to the disk scale, which makes it more difficult to magnetically align the grains on the disk scale.
Specifically, to produce the observed ${\sim} 1\%$ polarization at ${\sim} 100\,\mathrm{au}$ scale via grains aligned with the B-field, even relatively small grains of $1\,\mathrm{\mu m}$ in size need to have their magnetic susceptibilities significantly enhanced (by a factor of ${\sim} 20$) over the standard value, potentially through superparamagnetic inclusions. This requirement is more stringent for larger grains, with the enhancement factor increasing linearly with the grain size, reaching ${\sim} 2\times 10^4$ for millimeter-sized grains. Even if the required enhancement can be achieved, the resulting inferred magnetic field orientation in the simulation does not show a preference for the major axis, which is inconsistent with the observed pattern. We thus conclude that the observed trends are best described by the model where the polarization on the envelope scale is dominated by magnetically aligned grains and that on the disk scale by scattering.
\end{abstract}




\section{Introduction}

Magnetic fields have long been viewed as a key ingredient in molecular clouds and star formation \citep{MO07}. This view has been greatly strengthened by polarimetric observations of dust continuum emission in recent years, especially through the {\it Planck} all-sky survey \citep[e.g.,][]{PlanckXIX} and Atacama Large Millimeter/submillimeter Array \citep[ALMA, e.g.,][]{Hull_Zhang_2019}. On the relatively large cloud scales probed by {\it Planck}, there is little doubt that the polarized dust emission traces the magnetic field because the (relatively small, sub-micron-sized) dust grains are known to be preferentially aligned with their short axis along the magnetic field \citep[e.g.,][]{Andersson_2015}. On the much smaller scales probed by ALMA, the situation is less clear, which is the focus of our investigation.

In particular, ALMA has detected dust polarization in an increasing number of protostellar systems that are still deeply embedded in their massive envelopes \citep[e.g.,][]{Hull+17,Hull+20,Cox18,Maury+18,Sadavoy+18a,Sadavoy+18b,Sadavoy19,Kwon+19,LeGouellec+19,TakahashiS+19,KoCL+20,Yen+20}.
\citet{LeGouellec2020} analyzed the polarization data from Class 0 sources statistically, and found an interesting anti-correlation between the dispersion of the polarization orientations and the polarization fraction. Another interesting trend, first discussed in \citet{Cox18} and quantified further in this paper (see Section~\ref{sec:ALMAdata} below), is that the typical dust polarization fraction on the $10^3\,\mathrm{au}$ scale of the inner protostellar envelope (${\gtrsim} 5\%$) is much higher than that on the $10^2\,\mathrm{au}$ disk scale (${\lesssim} 1\%$). While the relatively high polarization fraction on the envelope scale is likely still due to magnetically aligned grains (as on the larger cloud scale), the origin of the much lower polarization fraction on the disk scale is less certain.

One possible explanation of the lower polarization fraction on the disk scale compared to that on the envelope scale is that the grains in the disk are much larger than those in the envelope and thus harder to align magnetically because of their longer Larmor precession timescale around the magnetic field (e.g.,~\citealt{Lazarian2007}; \citealt{Yang21}). The higher density in the disk also makes it harder for grain alignment, regardless of the alignment mechanism because of more frequent randomizing collisions of the grains with their ambient gas. These lead to a larger ratio of the Larmor precession timescale to the gas damping timescale in the disk compared to that in the envelope, and hence a lower magnetic grain alignment efficiency in the disk (see equation~\ref{eq:aligncrit} below), which may explain its lower polarization fraction.

A potential scenario is that the reduction factor of magnetic alignment on the disk scale is so large as to render the polarized emission from magnetically aligned grains undetectable. In this case, another mechanism is needed to explain the lower but still well-measured polarization at the ${\sim} 1\%$ level. The most likely alternative to magnetically aligned grains is dust self-scattering, which has been shown to be capable of producing percent-level polarization at (sub)millimeter wavelengths \citep{Kataoka15}, especially in inclined disks \citep{Yang16}.

The goal of our investigation is therefore to quantify the change in the dust polarization fraction and pattern in the transition from the protostellar envelope
scale to the disk scale and to determine whether the observed trends in the transition can be explained by magnetically aligned grains alone or whether scattering is also needed.
In particular, we examine how the condition for magnetic grain alignment at a given location in the protostellar system depends on its local environment, including the gas density, magnetic field strength, and the grain properties such as size and magnetic susceptibility. We also take into consideration the possible enhancement of dust magnetic susceptibility by superparamagnetic inclusions (SPIs) based on the theory recently developed by \citet{Yang21}. Our analytic model is then tested in synthetic polarization of a simulated protostellar disk-envelope system from \citet{Lam2019} to provide a qualitative comparison to the ALMA observations discussed in \citet{Cox18}.

The rest of the paper is organized as follows. We first revisit the ALMA polarization data of 8 protostellar systems in the Perseus molecular clouds presented in \citet{Cox18} to quantify the sharp drop in polarization fraction and change in polarization orientation from the protostellar envelope scale to the disk scale (Section~\ref{sec:ALMAdata}).
In Section~\ref{sec:theory}, we provide an overview of the condition for magnetic alignment of dust grains based on \citet{Yang21}.
The non-ideal MHD disk formation simulations of \citet{Lam2019} are described in Section~\ref{sec:method}, where we also present the synthetic polarization observations and their comparison with the ALMA data. Our results are discussed in Section~\ref{sec:discussion}.
Section~\ref{sec:summary} summarizes our conclusions.

\section{Observational Motivation: Sharp Envelope-Disk Transition in Dust Polarization}
\label{sec:ALMAdata}

\begin{figure*}
    \centering
    \includegraphics[width=\textwidth]{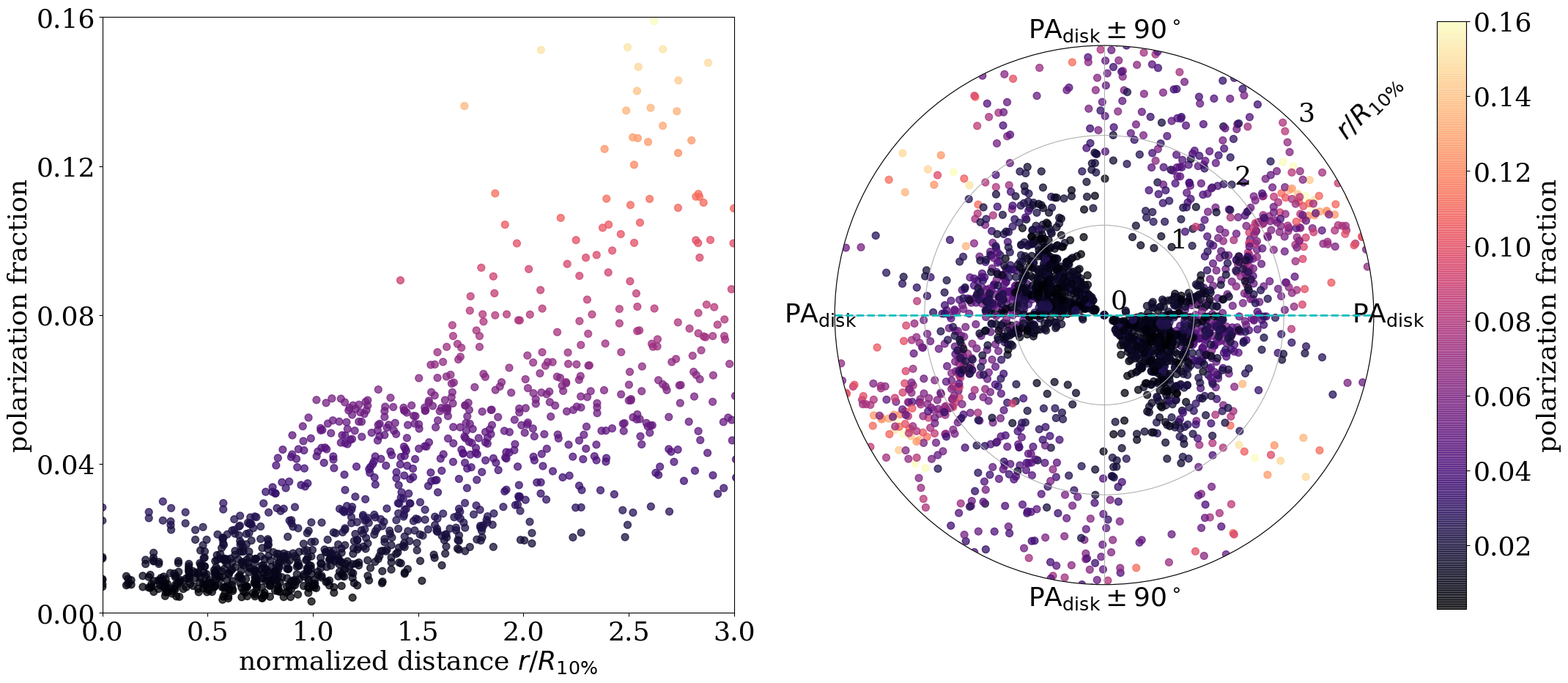}
    \caption{Trends of observed polarization from envelope to disk scales. Plotted as functions of the deprojected distance from the center (normalized by $R_{\rm 10\%}$; see text for definition) are the combined distributions of the polarization fraction ({\it left}) and the orientation of the polarization B-vector (rotated from the E-vector by 90$^\circ$; {\it right}), for 8 of the 10 protostellar systems presented in \citet{Cox18}.
    The orientation of the polar plot is rotated so that the disk major axes align with the horizontal line. Also, the polar plot is mirror-imaged to reflect the degeneracy of polarization angles over $180^\circ$. The polarization fraction is color coded in both panels to highlight its increase from the disk to envelope scale.
    }
    \label{fig:ObsTrends}
\end{figure*}

The combination of sensitivity and resolution afforded by ALMA makes it possible to detect polarized dust emission from the inner protostellar envelope down to the disk scale. The behavior of the polarization in the envelope-to-disk transition region is well illustrated in \citet{Cox18}, which presented ALMA Band 7 polarization data on 10 Class 0/I protostars in the Perseus molecular clouds (distance ${\sim}300\,\mathrm{pc}$; \citealt{Zucker19}). They found that on the 100-au disk scale the polarization tends to be well ordered, with a typical fraction of ${\lesssim} 1\%$. In contrast, on the larger envelope scale, the polarization becomes more disordered, with a relatively large fraction of typically ${\gtrsim} 5\%$. These differences were clearly visible in their polarization maps (see their Figs.~1-3) and quantified through cumulative probability distributions of the polarization fraction (see their Fig.~4).

To characterize the polarization in the envelope-disk transition further and facilitate a quantitative comparison of the observational data with synthetic observations based on numerical simulation, we re-examine the polarization data from \citet{Cox18} from a different angle by taking into account the spatial information more directly. In particular, we want to know how the polarization fraction and direction change with the distance from the center.
Since the resolution of the ALMA data presented in \citet{Cox18} is $\approx 0.4^{\prime\prime}$, we regridded the data to have pixel size $\approx 0.2^{\prime\prime}$ so that there are $\sim 2-4$ pixels per beam.
To correct for projection effects, we need to know the system orientation in the plane of the sky and inclination relative to the sky plane. For the orientation, we use the outflow directions inferred in \citet{2015ApJ...805..125T} and \citet{Stephens17,MASSES18}. Since protostellar outflows are generally aligned perpendicular to the equatorial plane of the system, we adopt the orientation $90^\circ$ to the outflow direction as the position angle of the disk of the system. For the inclination angle of the system,
we use the disk orientation inferred from the VLA Ka-band continuum observations by \citet{SeguraCox18}, as part of the VANDAM survey \citep{Tobin16}, whenever such measurement is available. For the other 5 systems that do not have a well-measured inclination, we assume $45^\circ$ for the inclination angle.\footnote{We have tested that the inclination angle does not have a huge effect (unless the system is nearly edge-on, which is less likely) on the features we show in Fig.~\ref{fig:ObsTrends}, since these features are visible even before de-projection (see Appendix~\ref{app:reproj}). We therefore adopt a representative value of $45^\circ$.}
Considering the targets are different in size, we further normalized the derived radius by a characteristic ``disk'' radius, defined as the radius of a circle that encloses the same area as the ellipse at 10\% maximum intensity from 2D Gaussian fitting of the ALMA Band 7 Stokes I image;
i.e., we fit the emission with 2D Gaussian distribution, measure the parameters of the ellipse at the 10\% peak level of the fitted distribution,
and calculate $R_{\rm 10\%} \equiv \sqrt{a\cdot b}$ where $a$, $b$ are the half lengths of the major and minor axes of the resulting ellipse from the fitted Gaussian, respectively.
The fitted Gaussian peak is adopted as the center of the system. The same masking criteria as in \citet{Cox18} are used, i.e., $I > 5\sigma_I$ and $P > 3\sigma_P$. More details on the de-projection process, as well as the parameters adopted for individual sources, are described in Appendix~\ref{app:reproj}.

In Fig.~\ref{fig:ObsTrends} we plot, as scatter plots color-coded by polarization fraction, the combined distributions of the polarization fraction (left panel) and the inferred orientation of the polarization B-vector (rotated from the E-vector by 90$^\circ$; right panel), as a function of the normalized (deprojected) distance from the center for 8 of the 10 sources discussed in \citet{Cox18}. We note that, among the 10 sources presented in \citet{Cox18}, Per\,41 has too few polarization detections for meaningful analysis. The other source, Per\,21, is located within the highly perturbed protostellar envelope NGC\,1333 IRAS\,7 that harbors at least two other protostars \citep[see e.g.,][]{Tobin16,ChenGBT19} with complex outflow morphologies \citep{Stephens17}. The complexity makes it difficult to assign an outflow direction for Per\,21 with confidence. We thus exclude it in this study as well.

The left panel of Fig.~\ref{fig:ObsTrends} shows that there is a clear trend for the polarization fraction to stay roughly constant at typically a percent level (although it is higher in a couple of cases; see discussion in Section~4) within about one normalized radius (i.e., $r/R_{\rm 10\%}\lesssim 1$) and then rapidly increases outwards, reaching values as high as $10-15\%$ or more. As shown in Appendix~\ref{app:reproj}, the trend is clear in both the cases where a large number of polarization vectors are detected on the envelope scale (such as Per\,2, Per\,5, Per\,11, and Per\,29) and those with fewer envelope-scale detections (such as Per\,14, Per\,18, and Per\,50). It is a robust feature that needs to be explained by all models of envelope/disk polarization, including our own (see Section~\ref{sec:method} below).

The right panel of Fig.~\ref{fig:ObsTrends} shows that the orientations of the polarization B-vectors are clearly non-isotropic (e.g., two cavities devoid of data points) within about one normalized radius ($R_{\rm 10\%}$), with a strong preference towards the disk orientation. This is in strong contrast with the orientations on the larger (envelope) scale, which are much more isotropic. The difference can also be seen in individual sources (e.g., the right panels of Fig.~\ref{fig:obsprhist}), which tend to have much more uniform polarization orientations (i.e., narrow distributions) on the disk scale than on the envelope scale.
Nevertheless, we note that there are noticeable offsets between the disk-scale polarization orientation and the outflow-based system major axis in some of our targets (Per\,5, Per\,11, Per\,18, and Per\,26; see Fig.~\ref{fig:obsprhist}). It is unclear what is the physical reason introducing such offset, and we discuss this further in Sec.~\ref{sec:discussion}.
In any case, the difference in the polarization orientations between the envelope and disk scales is significant, and is another feature that needs to be explained.

\section{Condition for Magnetic Grain Alignment}
\label{sec:theory}

The simplest explanation of the observed drop in polarization fraction from the envelope scale to the disk scale is that the grains are less well aligned at higher densities, especially inside the disks. Here, to explore this possibility quantitatively, we discuss the necessary conditions for magnetic grain alignment based on the recent work by \citet{Yang21}.

The investigation of magnetic alignment of spinning dust grains has a long and distinguished history (see \citealt{Andersson_2015} for a review). Although there are different proposed mechanisms to spin up the dust grains, the current favorite is radiative alignment torque, which appears capable of spinning up grains in protoplanetary disks \citep[e.g.,][]{Tazaki+17,Lazarian2007}. However, for the spinning grains to align with the magnetic field, they have to gyrate quickly around the magnetic field, with a Larmor precession timescale $t_L$ shorter than the gas damping timescale on dust grains $t_d$.

\subsection{Larmor Precession Timescale}

It is well-known that a spinning grain would be magnetized through the \citet{Barnett1915} effect.
This spinning-induced magnetization can be quantified as $\mathbf{M}=\chi \mathbf{\Omega}/\gamma$ \citep{P79,Roberge1993}, where $\chi$ is the magnetic susceptibility, $\mathbf{\Omega}$ is the angular velocity, \mbox{$\gamma = g\mu_B/\hbar$} is the gyromagnetic ratio with $\mu_B = 9.27 \times 10^{-21}$\,erg$\cdot$G$^{-1}$ (the Bohr magneton) and $g \approx 2$ (the $g$-factor) for electrons \citep{Draine1996}.
For a dust grain with a magnetic moment $|\mathbf{M}|V$, where $V$ is the volume of the dust grain (i.e.,~$V \equiv 4\pi a^3/3$ for spherical grains with radius $a$), the magnetic torque exerted on the dust grain by the external magnetic field $B$ is roughly $|\mathbf{M}|VB$.
We can therefore define a Larmor precession timescale as \citep[see e.g.,]{Lazarian2007,Yang21}:
\begin{equation}
    \begin{split}
    t_L =& \frac{2\pi I |\mathbf{\Omega}|} {|\mathbf{M}|VB} = \frac{4\pi(\rho_s V) a^2/5 \cdot \gamma}{\chi V B} =
    \left(\frac{4\pi\gamma}{5}\right)\frac{\rho_s a^2}{\chi B}\\
    =& \ 2.6\times 10^{11}\,\mathrm{s} \\
    &\times \hat{\chi}^{-1}
    \left(\frac{\rho_s}{3~\mathrm{g/cm^3}}\right)
    \left( \frac{T_s}{15~\mathrm{K}} \right) \left( \frac{B}{5~\mathrm{mG}} \right)^{-1} \left( \frac{a}{1~\mathrm{mm}} \right)^2,
    \end{split}
    \label{eq:tL}
\end{equation}
where $\rho_s$ is the mass density of the dust grain, and
\begin{equation}
\hat{\chi} \equiv \chi \cdot 10^3 \left( \frac{T_s}{15~\mathrm{K}}\right)
\end{equation}
is a dimensionless parameter of the magnetic susceptibility determined by the composition of dust grains, and $T_s$ is the dust temperature.

For regular paramagnetic material, the magnetic susceptibility is given as \citep[e.g.,][]{Morrish1980,Draine1996}
\begin{equation}
\chi_{\rm p} = \frac{n_{\rm p} \mu^2}{3kT},
\label{eq:chiC}
\end{equation}
where $n_{\rm p} = f_{\rm p} n_{\rm tot}$ is the number density of paramagnetic atoms with $f_{\rm p}$ the fraction of the atoms that are paramagnetic and $n_{\rm tot}$ the total atomic density, and $\mu = p \mu_B$ is the averaged Bohr magneton per iron atom.
\citet{Draine1996} noted that $p \approx 5.5$ for paramagnetic materials, and gave a realistic estimate of paramagnetic susceptibility:
\begin{equation}
    \chi_{\rm p} = 4.2 \times 10^{-2} f_{\rm p} \left(\frac{n_{\rm tot}}{10^{23}~\mathrm{cm^{-3}}}\right)
    \left(\frac{T_s}{15~\mathrm{K}}\right)^{-1} \left(\frac{p}{5.5}\right)^2.
\end{equation}
For typical interstellar dust grains composed of \{C, H, Mg, Si, Fe\}, $\rho_s \sim 3~{\rm g/cm}^3$ and $n_{\rm tot} \sim 10^{23}~{\rm cm}^{-3}$ \citep[see e.g.,][]{Draine1996}.
With $f_p\approx 0.1$ \citep{Draine1996}, this suggests that the dimensionless parameter $\hat\chi\sim 1$ in equation~(\ref{eq:tL}), which we will refer to as the ``standard" value for paramagnetic grains.

\subsection{Gas Damping Timescale}

Collisions between dust grains and surrounding gas particles have the potential to randomize the angular momenta of grains and thus limit the degree of magnetic alignment.
In general, the damping timescale can be estimated as the time needed to accumulate the same amount of mass from the gas material as the mass of the dust grain \citep[e.g.,][]{PS1971}.
Assuming every colliding gas particle sticks long enough on the grain surface for its kinetic energy to become thermalized at the grain temperature,
\citet{Roberge1993} derived a gas damping timescale
\begin{equation}
    \label{eq:td}
    \begin{split}
    t_d =& \frac{2\sqrt{\pi}}{5}\frac{\rho_s a}
    {n_{\rm g}m_{\rm g} {\mathrm{v}}_{\rm g,th}}\\
    =& 3.37\times 10^{8}~\mathrm{s}\\
    &\times \left(\frac{\rho_s}{3~\mathrm{g/cm^3}}\right)\left(\frac{a}{1~\mathrm{mm}}\right)\left(\frac{n_{\rm g}}{5\times10^9~\mathrm{cm^{-3}}}\right)^{-1}\left(\frac{T_{\rm g}}{15~\mathrm{K}}\right)^{-1/2}
    \end{split}
\end{equation}
for the simplest case of spherical grains\footnote{Note that for polarization produced by B-field aligned grains, non-spherical grains are required. Nonetheless, the timescale would be at the same order of magnitude.}, where $\rho_s$ and $a$ are the mass density and size of the grain, $n_\mathrm{g}$ the number density of the gas, $m_\mathrm{g}$ the mass per gas particle, and $\mathrm{v}_\mathrm{g,th}$ and $T_\mathrm{g}$ the gas thermal velocity and temperature, respectively.

\subsection{Condition for Magnetic Alignment with Superparamagnetic Inclusions}
\label{subsec:SPI}

Combining equations~(\ref{eq:tL}) and (\ref{eq:td}) and assuming that the dust grains and gas particles are in thermal equilibrium ($T_s = T_{\rm g} \equiv T$), we have
\begin{equation}
    \begin{split}
    \frac{t_L}{t_d} = 771 \times \left(\frac{a_{\rm mm}}{\hat\chi}\right) \left( \frac{B}{5~\mathrm{m G}} \right)^{-1} \left(\frac{n_{\rm g}}{5\times 10^9~\mathrm{cm^{-3}}}\right)  \left(\frac{T}{15~\mathrm{K}}\right)^{3/2}
    \label{eq:ratio}
    \end{split}
\end{equation}
where $a_{\rm mm} \equiv (a/1\,{\rm mm})$ is the grain size in units of millimeter.
Note that, as discussed in \citet{Yang21}, the magnetic susceptibility parameter $\hat\chi$ and the grain size $a$ are degenerate in the ratio between Larmor precession and gas damping.
Equation~(\ref{eq:ratio}) suggests that, for paramagnetic materials with the standard value of magnetic susceptibility $\hat\chi\sim 1$, only small, sub-micron-sized grains have $t_L < t_d$ (and thus can align with the magnetic field) for the adopted fiducial disk parameters of $B \sim 5~{\rm mG}$, $n_\mathrm{g}\sim 5\times10^9~{\rm cm}^{-3}$, and $T\sim 15~{\rm K}$. However, $\hat\chi$ can be enhanced by a large factor when superparamagnetic inclusions (SPIs hereafter) are present in the grains \citep{JS1967}.

Superparamagnetism appears in nanoparticles made of ferromagnetic or ferrimagnetic materials. Unlike paramagnetic materials that have only un-correlated electron spins,
within one superparamagnetic particle, all the atoms are spontaneously magnetized and behave like a single large magnetic moment (``macro-spin''; \citealt{BL1959}), which could greatly increase the magnetic susceptibility. \citet{Yang21} considered three types of candidate materials for the superparamagnetic inclusions, and estimated a maximum enhancement factor of ${\hat\chi}_{\rm max}\sim 1.1\times10^3$ for Fe$_3$O$_4$ (magnetite), $\sim 3.7\times 10^3$ for $\gamma$--Fe$_2$O$_3$ (maghemite), and $\sim 7.0\times 10^4$ for the extreme case of pure metallic iron (see their Table~1). Since magnetite and maghemite are more likely representative of superparamagnetic materials than metallic iron, the magnetic susceptibility is likely enhanced by a maximum factor up to a few thousands at most, rather than tens of thousands. \citet{Yang21} also noted that this maximum enhancement is a hard limit determined by the crystalline structure of the SPI material and the energy needed to overcome the crystalline structure and align the initially randomly oriented magnetic moments inside the SPIs to the direction of the external magnetic field. The required energy is proportional to the volume of the SPI. It therefore takes longer to align larger SPIs because of a larger energy required. Indeed, the alignment timescale is exponentially sensitive to the SPI size (see equation~9 of \citet{Yang21}), which severely limits the contributions of SPIs larger than the critical size to the magnetic susceptibility of the dust grain. Indeed, the maximum enhancement is reached only when all SPIs inside a grain have the (same) critical size, which is unlikely.

As briefly discussed in \citet{Yang21}, while no magnetic alignment is expected when the Larmor precession timescale is longer than the gas damping time scale ($t_L > t_d$), $t_L < t_d$ does not guarantee magnetic alignment. The grain needs to gyrate around the magnetic field multiple times within a single gas damping time in order to ensure magnetic alignment, which leads to a more stringent alignment condition: $t_L < t_d/\eta$, where $\eta (>1)$ is the number of gyrations per gas damping time needed for grain alignment. The exact value of $\eta$ is uncertain; in the discussion below, we will follow \citet{Yang21} and adopt a fiducial value of $10$. Making use of equation~(\ref{eq:ratio}), the alignment condition becomes:
\begin{equation}
    \label{eq:aligncrit}
    \lambda \equiv 771 \times
                   \left( \frac{B}{5\,\mathrm{mG}} \right)^{-1}
                   \left( \frac{n_\mathrm{g}}{5\times10^9\,\mathrm{cm^{-3}}} \right)
                   \left( \frac{T}{15\,\mathrm{K}} \right)^{3/2} < \xi
            \equiv \frac{\hat\chi/\eta}{a_\mathrm{mm}},
\end{equation}
where the dimensionless parameter $\lambda$ and $\xi$ encapsulate, respectively, the combination of the gas and magnetic field quantities and of the grain properties that enter the alignment condition. They will be referred to as gas and grain alignment parameter respectively hereafter. In regions where the gas alignment parameter $\lambda$ is larger, the grains are harder to magnetically align because of a higher gas density $n_{\rm g}$, a higher temperature $T$, or a lower magnetic field strength $B$. Conversely, grains with a larger grain alignment parameter $\xi$ are easier to align magnetically because of a larger magnetic susceptibility enhancement ${\hat \chi}$ by SPIs, a smaller grain size $a$, or a less stringent requirement on the number of times $\eta$ that the spinning grains need to gyrate around the field line before being knocked off by gas collisions in order to be magnetically aligned. In particular, in the protostellar envelope where the gas density $n_{\rm g}$ is lower and the grain size $a$ is expected to be smaller, the condition for magnetic alignment should be satisfied more easily. In the next section, we will quantify how the condition affects the polarization in the transition region from the protostellar envelope scale to the disk scale, using the physical quantities obtained from a non-ideal MHD disk formation simulation.

\section{Modeling Dust Polarization}
\label{sec:method}

\subsection{Model Setup and Synthetic Observations}
\label{sec:setup}

\begin{figure}
    \centering
    \includegraphics[width=\columnwidth]{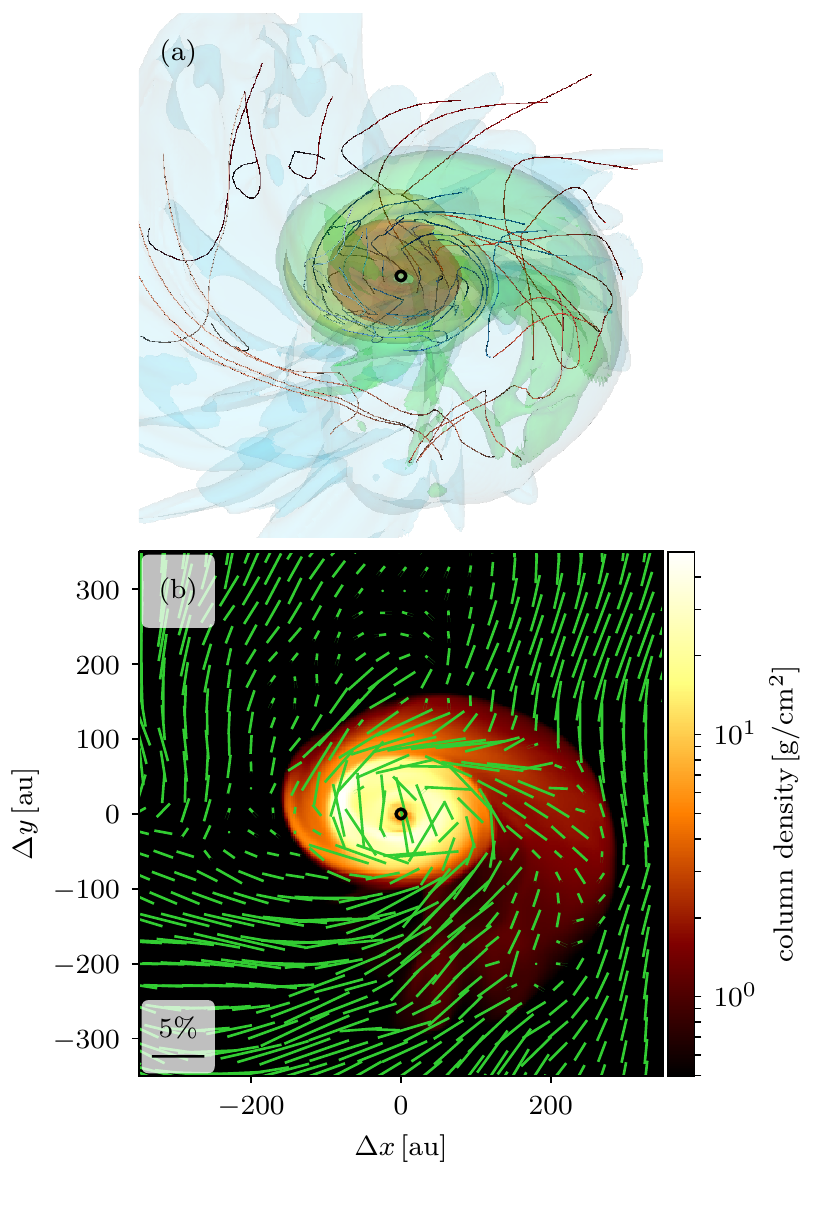}
    \caption{Panel (a): A 3D view of the density distribution and magnetic field lines of the simulated protostellar envelope-disk system that is used for the dust polarization modeling. Plotted are the iso-density surfaces at $n_\mathrm{g}=10^7$, $10^8$, $10^9$, $10^{10}\,\mathrm{cm^{-3}}$ (semi-transparent surfaces) and a sample of magnetic field lines, viewed at an angle of $45^\circ$ to the rotation axis (as in the synthetic polarization maps in the bottom panel and in Fig.~\ref{fig:chicomp} below). Panel (b): Polarization (B-)vectors (with length proportional to the polarization fraction) superposed on the color map of (mass) column density.}
    \label{fig:3D_view}
\end{figure}

The simulation to be used for our polarization modeling comes from \citet{Lam2019}, which contains a series of non-ideal MHD simulations of turbulent core collapse and disk formation with a range of values for the turbulent level and ambipolar diffusion (see their Table~1). The simulations are isothermal with a temperature of 10~K. They start with a centrally condensed core of $0.5~\mathrm{M_\odot}$ in mass and 2000~au in radius and an initial solid-body rotation of $\Omega=6\times 10^{-13}$~s$^{-1}$. For the purpose of illustrating the difference in dust polarization between the protostellar envelope and disk, we pick a high-resolution version of their model M1.0AD10.0, which has an initial turbulence Mach number of 1.0, a relatively large ambipolar diffusion coefficient that is ten times the fiducial value based on the standard cosmic ray ionization rate of $10^{-17}$~s$^{-1}$ \citep[see e.g.,][]{Shu1992}, and a ${\sim}100\,\mathrm{au}$ disk.
The simulation was performed using the {\tt Athena} code \citep{Athena2008} on a uniform base grid of 512$^3$. It was zoomed in once with half box length while keeping the number of cells fixed at 512$^3$, which yields a minimum resolution of $5$~au. The high-resolution simulation data at the time when the disk is well formed around the central protostar of $0.22~\mathrm{M_\odot}$ is used for our polarization analysis. The disk can be clearly see in the top panel of Fig.~\ref{fig:3D_view}, which shows a 3D view of the density distribution, together with several representative magnetic field lines. The bottom panel shows the polarization vectors obtained through the procedure discussed next.

In the simplest case of optically thin dust with a spatially homogeneous grain alignment efficiency, the Stokes parameters $I$, $Q$, and $U$ of dust thermal emission along a given line of sight are given by \citep[see e.g.,][]{fiegepudritz2000}:
\begin{subequations}
\begin{align}
    I &= \int \rho \left( 1 - \alpha \left( \frac{\cos^2 \gamma}{2} - \frac{1}{3} \right) \right) ds,\\
    Q &= \alpha \int \rho\,\cos 2\psi\,\cos^2 \gamma\,ds,\\
    U &= \alpha \int \rho\,\sin 2\psi\,\cos^2 \gamma\,ds,
\end{align}
    \label{eq:synpolideal}
\end{subequations}
where $\rho$ and $s$ are, respectively, the mass density at a given location and the distance into the cloud of that location along the line of sight. The polarizability parameter $\alpha$ is determined by the grain cross sections and alignment properties and assumed to be spatially constant. The quantity $\gamma$ is the inclination angle of the field line with respect to the plane of the sky, and $\psi$ is the angle of the magnetic field from the direction of positive $Q$ in the sky plane. The maximum degree of polarization $p_0$ is related to the parameter $\alpha$ through $p_0 = \alpha / (1 - \alpha/6)$. For simplicity, we set $\alpha=10\%$, which yields $p_0=10.17\%$\footnote{The maximum polarization parameter $p_0$ should in principle be determined from grain alignment theory, as done in, e.g.,~\citet{Valdivia+19,Kuffmeier+20}, using the POLARIS code \citep{POLARIS}. However, the uncertainties in the grain properties, especially their shapes, make it difficult to firmly predict this parameter. Our choice is guided by the typical values observed in the inner protostellar envelopes of the Perseus protostars. It is the same as the constant value adopted by \citet{Valdivia+19} but somewhat smaller than the value of $15\%$ adopted by \citet{Padovani+12} and \citet{Lee_Hull+17}.}.
The polarization fraction and direction are obtained from the Stokes parameters through
\begin{equation}
    p = \frac{\sqrt{Q^2 + U^2}}{I},\ \ \
    \phi = \frac{1}{2}{\rm arctan2}(U,Q).
\end{equation}

There are a few complications. First, the densest part of the protostellar disk formed in our simulation is moderately optically thick, which could lower its polarization fraction somewhat \citep{Yang+17,Lin+20a}. The optical depth (and associated extinction) is accounted for by solving the vector radiation transfer equation, as described in Appendix \ref{app:extinction}. Second, the magnetic alignment efficiency is not expected to be spatially homogeneous, with grains in denser regions less likely magnetically aligned, as discussed in Section~\ref{sec:theory}. We capture this effect using a magnetic alignment probability $A$, defined in each voxel of the simulation based on the alignment condition, equation~(\ref{eq:aligncrit}):
\begin{equation}
    A =
    \begin{cases}
        1, & \text{if the voxel satisfies equation~(\ref{eq:aligncrit})} \\
        0, & \text{if the voxel does not satisfy equation~(\ref{eq:aligncrit})}.
    \end{cases}
    \label{eqn:frac}
\end{equation}

Finally, it is well known that dust scattering can provide significant polarization in inclined disks along the disk minor axis \citep{Kataoka15,Yang16}. To account for this possibility, in one of the models, we added a scattering-induced polarization of fraction $p_{\rm sca}$ (see Section~\ref{sec:results} below for discussions on the value of $p_{\rm sca}$) along the disk minor axis (which is the negative $Q$ direction in our setup) only in those voxels that do not satisfy the magnetic alignment condition (i.e., $A=0$).

The synthetic observations are conducted at a resolution of 5\,au. To provide better comparison to the ALMA observations of \citet{Cox18}, we smoothed the synthetic $I$, $Q$, and $U$ maps with a 2D Gaussian beam with FWHM = 10~pixels (corresponding to $\sim 50$\,au), which is large enough to illustrate the beam convolution effect but small enough that the envelope and disk scales remain distinct (see Fig.~\ref{fig:chicomp} below).
To be consistent with our analysis on the observational data, we downsampled the smoothed synthetic observations to have $\approx 4$ pixels per beam.
We did not include any noise in the synthetic observations but applied the same masking criteria ($I > 5\sigma_I$, $P > 3\sigma_P$, where $P=\sqrt{Q^2+U^2}$ is the polarized intensity and $\sigma_I$, $\sigma_P$ are mean values measured from a relatively quiescent $10\times 10$ pixel$^2$ region on our synthetic maps) as adopted in \citet{Cox18} for consistency and better comparison with data.

\subsection{Model Results}
\label{sec:results}

\begin{figure*}
    \centering
    \includegraphics[width=\textwidth]{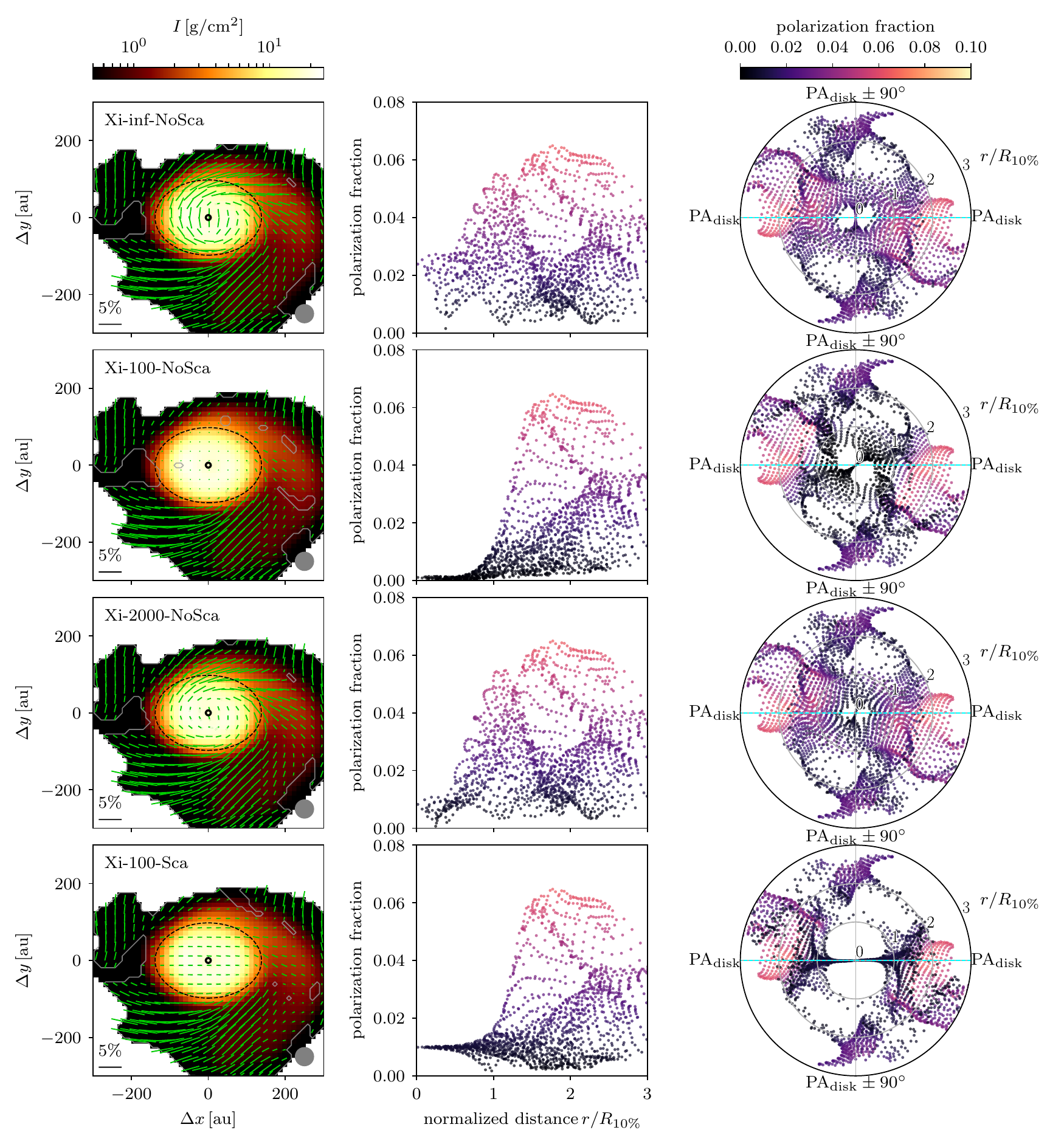}
    \caption{Comparing the 4 models of synthetic polarization considered in this study. Top to bottom: Model \mbox{Xi-inf-NoSca}, \mbox{Xi-100-NoSca}, \mbox{Xi-2000-NoSca}, and \mbox{Xi-100-Sca} (see Table~\ref{tab:parameters} for model parameters). {\it Left panels:} Maps of total intensity (with mask $I > 5 \sigma_I$), overplotted with polarization (B-)vectors {\it (green segments)}, with length proportional to the polarization fraction. Gray contours show the masking boundaries $P = 3\sigma_P$ adopted when calculating the scatter plots ({\it middle and right panels}).  Black open circles mark the locations of the protostar, and black dashed ellipses represent the disk defined by horizontal position angle and $45^\circ$ inclination, with size comparable to the contour of $10 \%$ peak intensity). {\it Middle and right panels:} scatter plots of polarization fraction ({\it middle}) and angle ({\it right}) in the corresponding envelope-disk system, as in Fig.~\ref{fig:ObsTrends}. The polarization fraction is color coded in both panels.}
    \label{fig:chicomp}
\end{figure*}

We consider 4 models that cover a range of parameters, with the simulated system viewed along a representative line of sight that is $45^\circ$ to the disk plane (i.e., an inclination angle $i=45^\circ$). The model names and parameters are listed in Table~\ref{tab:parameters}.
We follow the same deprojection process as described in Section~\ref{sec:ALMAdata} and Appendix~\ref{app:reproj} with the known projected disk orientation (horizontal) and inclination angle (45$^\circ$).
Similar to Fig.~\ref{fig:ObsTrends},
we plot the polarization fraction and angle as functions of the normalized, inclination-corrected radius for each synthetic observation in Fig.~\ref{fig:chicomp}.

\begin{table}
  \caption{Model parameters and outcome}
  \label{tab:parameters}
  \begin{tabular}{lccl}
    \hline
    \hline
    Model Name & $\xi$ & $p_\mathrm{sca}$ & Consistency\tablefootmark{a}  \\
    \hline
    \mbox{Xi-inf-NoSca}  & $\infty$ &  0.0 & no \\
    \mbox{Xi-100-NoSca}  & 100 & 0.0  & no  \\
    \mbox{Xi-2000-NoSca} & 2000 & 0.0 & partially  \\
    \mbox{Xi-100-Sca} & 100 & $1\%$  & yes \\
    \hline
  \end{tabular}
  \tablefoot{
    \tablefoottext{a}{Whether or not the model is broadly consistent with the observed polarization trends from the envelope to disk scale discussed in Section~\ref{sec:ALMAdata}.}
  }
\end{table}

We start the discussion with the simplest case of spatially homogeneous magnetic grain alignment without scattering (Model \mbox{Xi-inf-NoSca}, the first row of Fig.~\ref{fig:chicomp}). In this case, the alignment condition is satisfied everywhere, including in the densest part of the disk. To facilitate comparison with observations, we masked out regions with polarized intensity $P< 3 \sigma_P$, which creates several spatially coherent patches of high polarization separated by stripes of lower polarization. Since the dust is aligned with the magnetic field everywhere, the relatively-low polarization stripes are caused by a combination of field orientation along the line of sight and, more importantly, the variation of the field component in the sky plane, which leads to a cancellation of the polarized emission \citep[e.g.,][]{Kataoka+12}. A clear example of the lower polarization stripes is located to the upper-left of the disk (see the upper-left panel of Fig.~\ref{fig:chicomp} and, more clearly, Fig.~\ref{fig:3D_view}b, where the polarization map is not beam-convolved). The low polarization in this region comes from its magnetic field lines being pinching by the radial infall and twisted by rotation at the same time, which results in roughly orthogonal magnetic fields at different locations along the same line of sight (see  Fig.~\ref{fig:3D_view}a for a visual impression).

It is obvious that the simplest model fails to reproduce the observed trends in two ways. First, although the polarization fraction on the envelope scale is broadly consistent with the observed values, that on the disk scale is higher than typically observed. This is not too surprising since the magnetic field in the densest part of the disk that dominates the dust emission is rather well-ordered (and not along the line of sight; see the representative field lines threading the disk in Fig.~\ref{fig:3D_view}a for an illustration), which leads to relatively little cancellation of polarized emission. Indeed, the polarization fraction would be even higher without the beam-convolution, as can be seen from Fig.~\ref{fig:3D_view}a, which shows that the intrinsic polarization fraction (before beam convolution) on the disk scale is comparable to that on the envelope scale. Clearly, magnetic field geometry alone cannot explain the large reduction of the polarization fraction on the disk scale compared to the envelope scale. Second, there is substantial variation of the polarization orientations on the disk scale except in the very central region ($r\lesssim 0.4 R_{\rm 10\%}$), where the polarization is aligned perpendicular to the disk major axis.  The reason for this polarization orientation is that the magnetic field threading the disk has a significant poloidal component (as opposed to being wound up by disk rotation into a completely toroidal configuration, presumably because of the relatively large magnetic diffusivity that enabled the disk to form and survive in the simulation in the first place). We should note that the beam convolution makes the orientations of the polarization vectors in the inner part of the disk more ordered but this effect is relatively moderate (compare Figs.~\ref{fig:3D_view}b and the top-left panel of Fig.~\ref{fig:chicomp}). In particular, it does not make the polarization (B-)vectors preferentially align with the system major axis, which is one of the observed trends for the disk-scale polarization (see the right panel of Fig.~\ref{fig:ObsTrends}).

We next consider Model \mbox{Xi-100-NoSca} with the grain alignment parameter $\xi$ on the right side of the equation~(\ref{eq:aligncrit}) set to 100 instead of $\infty$. The results are shown in the second row of Fig.~\ref{fig:chicomp}. Note that the combination of physical parameters to reach $\xi\equiv \hat{\chi}/(\eta a_\mathrm{mm})=100$ is not unique. For the fiducial choice of $\eta=10$, the value corresponds to micron-sized  ($a_\mathrm{mm}=0.001$), regular paramagnetic grains without any superparamagnetic inclusions ($\hat{\chi}=1$). Another combination is to have much larger, mm-sized grains ($a_\mathrm{mm}=1$) with the magnetic susceptibility enhanced by a factor of 1000 ($\hat{\chi}=10^3$) by SPI. Compared with the simplest model with a spatially homogeneous grain alignment (Model \mbox{Xi-inf-NoSca}, first row), the polarization on the envelope scale appears little affected, indicating that the grains there remain efficiently aligned with the magnetic field. In contrast, the polarization on the disk scale is drastically reduced, to a level well below $1\%$.

\begin{figure*}
    \centering
      \includegraphics[width=\textwidth]{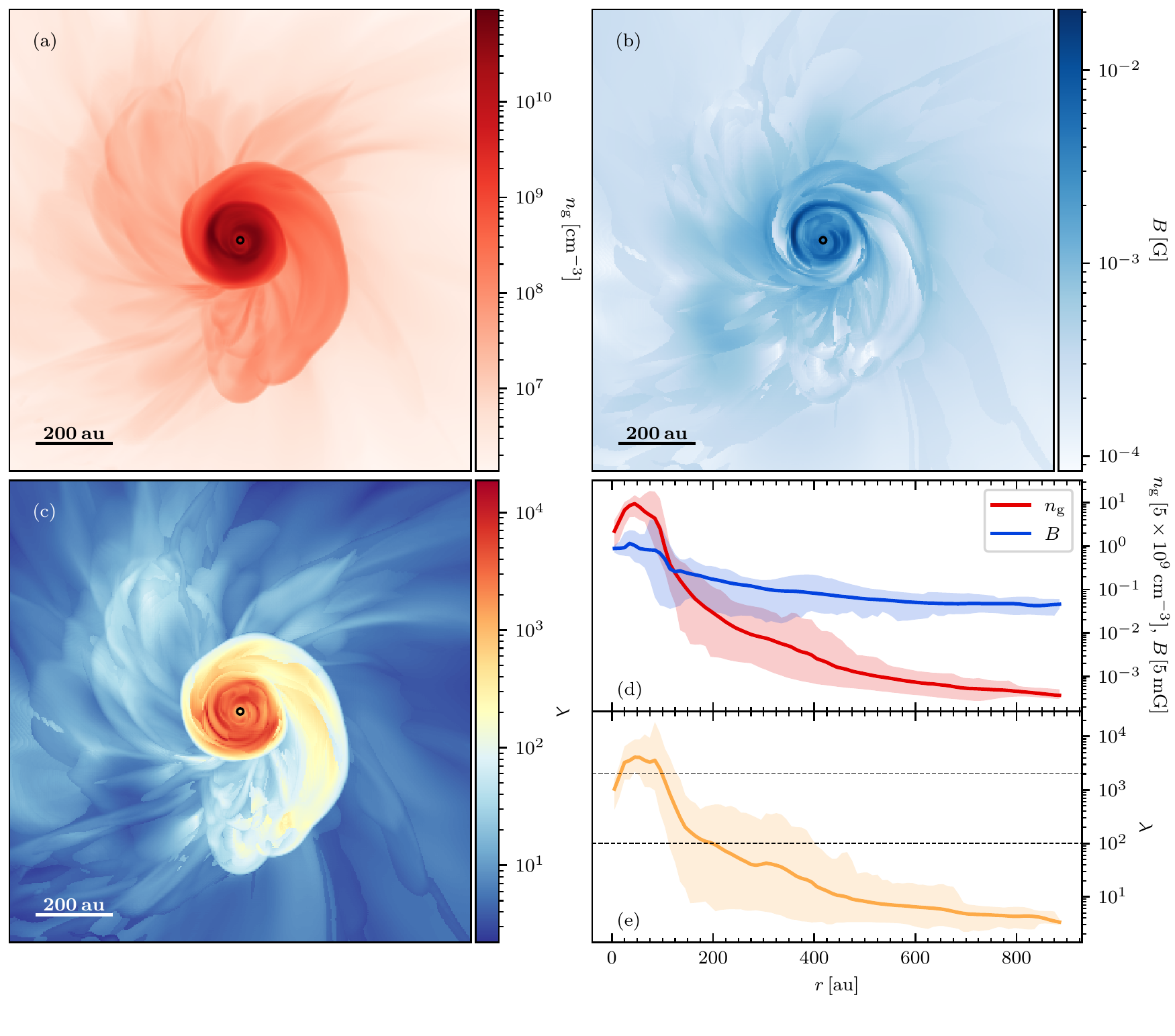}
    \caption{Face-on view of the spatial distributions (in the $x-y$ plane) of (a) volume density, (b) the magnetic field strength, and (c) the dimensionless gas alignment parameter $\lambda$  (see equation~\ref{eq:aligncrit}), at the location along each $z-$sight line where the density is the highest.  The small black circle highlights the location of the sink (stellar) particle in the simulation. Also plotted are the azimuthal average and range of the density and field strength (panel d) and gas alignment parameter $\lambda$ (panel e) as a function of radius. Note that $\lambda$ is larger in the disk than in the envelope, indicating that the condition for magnetic alignment is harder to satisfy in the former than in the latter. The grain alignment parameter adopted in Models \mbox{Xi-100-NoSca} and \mbox{Xi-100-Sca} ($\xi=100$) is shown as the lower dashed horizontal line in panel (e), and that in Model \mbox{Xi-2000-NoSca} ($\xi=2000$) as the upper dashed line.}
    \label{fig:lambda}
\end{figure*}

The reason for the reduction can be understood from Fig.~\ref{fig:lambda}, where we show a face-on view (along the $z-$axis) of the spatial distributions (in the $x-y$ plane) of the gas number density $n_\mathrm{g}$, the magnetic field strength $B$, and the corresponding dimensionless gas alignment parameter $\lambda$ defined in equation~(\ref{eq:aligncrit}). Along each $z-$sight line, these quantities are plotted at the location where the density is the highest, which is chosen to highlight the disk.
It is clear that, while both the magnetic field strength and the density increase from the envelope to the disk, the density increases by a much larger factor, making it easier for the collisions with gas particles to damp out the Larmor precession of the spinning grains around the magnetic field in the disk than in the envelope. This drastic difference between how the density and field strength vary from the envelope to the disk scale is quantified in panel (d), where we plot the average and range of these two quantities at each radius. The difference is reflected in the distribution of the gas alignment parameter $\lambda\propto n_{\rm g}/B$, which is plotted in panel (e). Clearly, $\lambda$ increases rapidly as the radius decreases, crossing the value of 100 adopted for the grain alignment parameter $\xi$ for Model \mbox{Xi-100-NoSca} (the lower dashed line in the panel) around a radius of order 200~au.
In the envelope outside this radius, the grain alignment condition $\lambda < \xi$ (equation~\ref{eq:aligncrit}) is satisfied, which leads to a high polarization fraction. Interior to this radius, the alignment condition is violated for most of the mass, which leads to a polarization fraction well below the observed value. An implication of this deficiency is that the observed disk-scale polarization is unlikely produced by large, mm-sized, magnetically aligned grains since it would require an unrealistically large enhancement of the magnetic susceptibility (by a factor more than $10^3$).

To increase the polarization level on the disk scale, we consider a larger grain alignment parameter, $\xi=2000$ (Model \mbox{Xi-2000-NoSca}, third row of Fig.~\ref{fig:chicomp}). For the fiducial value of $\eta=10$, this choice corresponds to $\hat{\chi}=20000\ a_\mathrm{mm}$, which means that, for large mm-sized grains, the magnetic susceptibility must be enhanced by a factor close the maximum possible value for the extreme case of pure metallic iron as the material for superparamagnetic inclusions \citep{Yang21}; the requirement would be even more extreme when the higher temperature on the disk scale is taken into account. For smaller grains, the requirement is less extreme. For example, for 1~$\mu$m-sized grains, the enhancement factor is 20. In any case, with the grain alignment parameter $\xi$ increased from 100 to 2000, we are able to increase the polarization fraction on the disk scale from $\ll 1\%$ to $\sim 1\%$, more in line with the typically observed values. At the same time, this polarization fraction is smaller than that of Model \mbox{Xi-inf-NoSca}, where the magnetically aligned grains in the disk produce a polarization fraction significantly above the typically observed values.

Even though Model \mbox{Xi-2000-NoSca} has enough magnetic alignment of the grains to produce a polarization fraction in line with the typically observed values on the disk scale, its polarization orientations differ substantially from the observed trend. In particular, there is a significant variation in polarization orientation on the disk scale, with a preference along the disk minor axis, which is the opposite of the observed trend. This discrepancy motivates us to consider a model that includes another polarization mechanism -- dust scattering.

The degree of continuum polarization produced by dust scattering depends sensitively on the grain size. At the wavelength of ALMA Band 7, the optimal size is of order $10^2\ \mu$m \citep{Kataoka15}. Such grains would typically produce a percent-level polarization with E-vectors along the minor axis of the inclined disk \citep[e.g.,][]{Kataoka+16,Yang16}. To capture this effect, we add in Model \mbox{Xi-100-Sca} a scattering-induced polarization to Model \mbox{Xi-100-NoSca} according to the prescription given in equation~(\ref{eq:scattering}), with $p_\mathrm{sca}=0.01$. The result is displayed in the last row of Fig.~\ref{fig:chicomp}.

As expected, the polarization structure of Model \mbox{Xi-100-Sca} is very similar to that of Model \mbox{Xi-100-NoSca} on the envelope scale, where the grains remain mostly aligned to the magnetic field in both cases (comparing the second and fourth row of Fig.~\ref{fig:chicomp}). The main difference comes from the disk scale, where the scattering in the former has now produced a percent-level polarization with the B-vectors preferentially along the major axis of the disk, as suggested by the ALMA observations in \citet{Cox18}. This hybrid model, with polarization on the envelope scale dominated by magnetically aligned grains and that on the disk scale by scattering, is thus the best of the four representative models considered in this paper for interpreting the observational results.

\section{Discussion}
\label{sec:discussion}

We would like to point out that since the simulations in \citet{Lam2019} are isothermal, the temperature dependence in the gas alignment parameter $\lambda$ (see equation~\ref{eq:aligncrit}) is neglected in our analysis in Section~\ref{sec:results}. However,
we note that the increase in $\lambda$ from the envelope to the disk scale is expected to be even faster when the temperature gradient is taken into account under the following considerations.
Generally speaking, the temperature and thermal velocity (proportional to $\sqrt{T}$) are expected to be higher at small radii. Assuming hydrostatic equilibrium, the thickness of the disk is proportional to the thermal velocity, and the gas density is inversely proportional to the disk height. This decrease in density as a result of the thicker disk cancels out the dependence of collision frequency (or the gas damping timescale; see equation~\ref{eq:td}) on the increasing temperature, resulting in a similar gas damping timescale. However, the higher temperature also reduces the magnetic susceptibility and increases the Larmor precession timescale (see equation~\ref{eq:tL}), making magnetic alignment more difficult for dust grains in the disk. Therefore, including a temperature dependence in our analysis in Section~\ref{sec:results} would not enhance magnetic alignment on the disk scale, and thus our qualitative conclusion would be strengthened.

Though we present the observational data from 8 sources as a combined plot in Fig.~\ref{fig:ObsTrends}, it is important to consider the differences among individual sources. In particular, we note that some of the protostellar systems have higher disk-scale polarization level than the others. This can be seen in the left panel of Fig.~\ref{fig:ObsTrends}: While the majority of the cells at $r/R_{\rm 10\%}\lesssim 0.5$ have polarization fraction $\approx 0.01$, there are systems showing $p\approx 0.02-0.03$ at the inner-most region of the disk (also see Fig.~\ref{fig:obsprhist}). The detailed analysis of individual protostellar systems is beyond the scope of this paper, but we would like to note the possibility that though dust scattering works well for the general trend of the disk-scale polarization, magnetically aligned grains may still dominate the polarization observed in some protostellar systems if the grains have relatively small sizes and/or their magnetic susceptibility is greatly enhanced by SPIs.

We also note that, while the preference of the B-vectors to align with the major axis of the disk is clear in the ALMA data, they could be offset by up to $\sim 45^\circ$ (see Sec.~\ref{sec:ALMAdata}; also see Fig.~\ref{fig:obsprhist}). Since we derived the direction of the disk major axis from the outflow orientation, which has a relatively low uncertainty (typically $\lesssim 10^\circ$; see e.g.,~\citealt{MASSES18}), this offset is likely real if the outflow is launched perpendicular to the disk, as generally expected.  It is possible that in the systems with large polarization-disk orientation offsets (Per\,5, Per\,11, Per\,18, and Per\,26; see Fig.~\ref{fig:obsprhist}), the scattering-induced polarization on the disk scale is contaminated by that from magnetically aligned grains, whose orientation can deviate significantly from the disk major axis (see, e.g., the top panel of Fig.~\ref{fig:chicomp}). Higher resolution observations are needed to test this possibility.

\section{Summary}
\label{sec:summary}

We have re-analyzed the ALMA Band 7 polarization data from \citet{Cox18} for deeply embedded protostars in the Perseus molecular cloud on the scales of inner protostellar envelopes and disks. A simple dust polarization model was constructed based on recent theoretical work by \citet{Yang21} and the non-ideal MHD disk formation simulations of \citet{Lam2019} to explain the observational trends. Our main results are summarized as follows:

1. Using scatter plots, we quantified the observational trends first identified in \citet{Cox18} that the polarization fraction stays relatively constant at a typical level of $\sim 1\%$ on the disk scale ($r \lesssim R_{10\%}$) and increases sharply going from the disk to the envelope scale (see the left panel of Fig.~\ref{fig:ObsTrends}). In addition, the polarization \mbox{(B-)vectors} tend to orient more perpendicular than parallel to outflow axis (right panel of Fig.~\ref{fig:ObsTrends}), indicating a preferential alignment with the disk major axis, under the assumption that all polarization is from magnetically aligned grains. These quantitative behaviors provide guidance to theoretical models of dust polarization in the earliest phases of low-mass star formation.

2. Using MHD simulations of disk formation enabled by a combination of ambipolar diffusion and turbulence \citep{Lam2019}, we showed that the observed sharp reduction of polarization fraction from the envelope to disk scale cannot be explained by the magnetic field geometry alone. The magnetic field on the disk scale is rather well ordered (see Fig.~\ref{fig:3D_view}) and produces an intrinsic polarization fraction comparable to that in the envelope. Beam averaging can significantly reduce the polarization fraction on the disk scale, making it more consistent with the typically observed values. It cannot, however, make the orientations of the polarization (B-)vectors preferentially align with the system major axis (see Fig.~\ref{fig:chicomp}, first row), which is one of the observed trends.

3. Our MHD simulations show that the magnetic field strength increases from the envelope to the disk scale (Fig.~\ref{fig:lambda}a,d), which tends to increase the ability of the magnetic field to align the spinning grains. However, this tendency is overwhelmed by the much faster increase in density (Fig.~\ref{fig:lambda}b,d), making grain alignment more difficult through more frequent gas collisions. We show that large, mm-sized grains cannot be aligned by the magnetic field on the disk scale to produce the typically observed polarization level of $\sim 1\%$ even if their magnetic susceptibility is enhanced by an uncomfortably large factor of $10^3$ (see the second row of Fig.~\ref{fig:chicomp}). In order for such large grains to be aligned well enough to account for the percent level polarization, the enhancement factor must reach a factor of $\sim 2\times 10^4$ (see the third row of Fig.~\ref{fig:chicomp}), which is unlikely since it is close to the maximum factor estimated for the extreme case of metallic iron as the material for superparamagnetic inclusions. For smaller grains, the required enhancement factor is less extreme. For example, for 1~$\mu$m grains, it is a factor of 20, which is still quite significant; it is unclear whether such an enhancement can be naturally achieved or not. Even if the required enhancement can be achieved, the resulting polarization (B-vector) orientation does not show a preference for the major axis on the disk scale, which is inconsistent with the observed trend.

4. The model most consistent with the observed trends is the one where the polarization on the envelope scale is dominated by magnetically aligned grains (with a relatively large maximum polarization fraction of order $10\%$) and by scattering on the disk scale. The former is consistent with the expectation that the grains in the low-density envelope remain relatively small and thus more easily aligned magnetically because of fast Larmor precession and long gas damping timescale. The latter requires grains of order $0.1$~mm in order to efficiently produce polarization at (sub)millimeter wavelengths through scattering.

\section*{Acknowledgements}

KHL acknowledges support from NRAO ALMA SOS awards and NSF AST-1716259. CYC is supported in part by VITA (Virginia Institute for Theoretical Astrophysics) and NSF AST-1815784. ZYL is supported in part by NASA 80NSSC20K0533 and 80NSSC18K1095. LWL acknowledges support from NSF AST-1910364. Resources supporting this work were provided by the NASA High-End Computing (HEC) Program through the NASA Advanced Supercomputing (NAS) Division at Ames Research Center based on NASA grant 80NSSC18K0481.

\section*{Data Availability}

The data underlying this article will be shared on reasonable request to the corresponding author.





\begin{thebibliography}{}
\makeatletter
\relax
\def\mn@urlcharsother{\let\do\@makeother \do\$\do\&\do\#\do\^\do\_\do\%\do\~}
\def\mn@doi{\begingroup\mn@urlcharsother \@ifnextchar [ {\mn@doi@}
  {\mn@doi@[]}}
\def\mn@doi@[#1]#2{\def\@tempa{#1}\ifx\@tempa\@empty \href
  {http://dx.doi.org/#2} {doi:#2}\else \href {http://dx.doi.org/#2} {#1}\fi
  \endgroup}
\def\mn@eprint#1#2{\mn@eprint@#1:#2::\@nil}
\def\mn@eprint@arXiv#1{\href {http://arxiv.org/abs/#1} {{\tt arXiv:#1}}}
\def\mn@eprint@dblp#1{\href {http://dblp.uni-trier.de/rec/bibtex/#1.xml}
  {dblp:#1}}
\def\mn@eprint@#1:#2:#3:#4\@nil{\def\@tempa {#1}\def\@tempb {#2}\def\@tempc
  {#3}\ifx \@tempc \@empty \let \@tempc \@tempb \let \@tempb \@tempa \fi \ifx
  \@tempb \@empty \def\@tempb {arXiv}\fi \@ifundefined
  {mn@eprint@\@tempb}{\@tempb:\@tempc}{\expandafter \expandafter \csname
  mn@eprint@\@tempb\endcsname \expandafter{\@tempc}}}

\bibitem[\protect\citeauthoryear{{Andersson}, {Lazarian}  \&
  {Vaillancourt}}{{Andersson} et~al.}{2015}]{Andersson_2015}
{Andersson} B.~G.,  {Lazarian} A.,   {Vaillancourt} J.~E.,  2015, \mn@doi
  [\araa] {10.1146/annurev-astro-082214-122414}, \href
  {https://ui.adsabs.harvard.edu/abs/2015ARA&A..53..501A} {53, 501}

\bibitem[\protect\citeauthoryear{{Barnett}}{{Barnett}}{1915}]{Barnett1915}
{Barnett} S.~J.,  1915, \mn@doi [Physical Review] {10.1103/PhysRev.6.239},
  \href {https://ui.adsabs.harvard.edu/abs/1915PhRv....6..239B} {6, 239}

\bibitem[\protect\citeauthoryear{{Bean} \& {Livingston}}{{Bean} \&
  {Livingston}}{1959}]{BL1959}
{Bean} C.~P.,  {Livingston} J.~D.,  1959, \mn@doi [Journal of Applied Physics]
  {10.1063/1.2185850}, \href
  {https://ui.adsabs.harvard.edu/abs/1959JAP....30S.120B} {30, S120}

\bibitem[\protect\citeauthoryear{{Chen} et~al.,}{{Chen}
  et~al.}{2019}]{ChenGBT19}
{Chen} C.-Y.,  et~al., 2019, \mn@doi [\mnras] {10.1093/mnras/stz2633}, \href
  {https://ui.adsabs.harvard.edu/abs/2019MNRAS.490..527C} {490, 527}

\bibitem[\protect\citeauthoryear{{Cox}, {Harris}, {Looney}, {Li}, {Yang},
  {Tobin}  \& {Stephens}}{{Cox} et~al.}{2018}]{Cox18}
{Cox} E.~G.,  {Harris} R.~J.,  {Looney} L.~W.,  {Li} Z.-Y.,  {Yang} H.,
  {Tobin} J.~J.,   {Stephens} I.,  2018, \mn@doi [\apj]
  {10.3847/1538-4357/aaacd2}, \href
  {https://ui.adsabs.harvard.edu/abs/2018ApJ...855...92C} {855, 92}

\bibitem[\protect\citeauthoryear{{Draine}}{{Draine}}{1996}]{Draine1996}
{Draine} B.~T.,  1996, in {Roberge} W.~G.,  {Whittet} D. C.~B.,  eds,
  Astronomical Society of the Pacific Conference Series Vol. 97, Polarimetry of
  the Interstellar Medium. p.~16 (\mn@eprint {arXiv} {astro-ph/9603053})

\bibitem[\protect\citeauthoryear{{Fiege} \& {Pudritz}}{{Fiege} \&
  {Pudritz}}{2000}]{fiegepudritz2000}
{Fiege} J.~D.,  {Pudritz} R.~E.,  2000, \mn@doi [\apj] {10.1086/317228}, \href
  {https://ui.adsabs.harvard.edu/abs/2000ApJ...544..830F} {544, 830}

\bibitem[\protect\citeauthoryear{{Hull} \& {Zhang}}{{Hull} \&
  {Zhang}}{2019}]{Hull_Zhang_2019}
{Hull} C. L.~H.,  {Zhang} Q.,  2019, \mn@doi [Frontiers in Astronomy and Space
  Sciences] {10.3389/fspas.2019.00003}, \href
  {https://ui.adsabs.harvard.edu/abs/2019FrASS...6....3H} {6, 3}

\bibitem[\protect\citeauthoryear{{Hull} et~al.,}{{Hull} et~al.}{2017}]{Hull+17}
{Hull} C. L.~H.,  et~al., 2017, \mn@doi [\apj] {10.3847/1538-4357/aa7fe9},
  \href {https://ui.adsabs.harvard.edu/abs/2017ApJ...847...92H} {847, 92}

\bibitem[\protect\citeauthoryear{{Hull}, {Le Gouellec}, {Girart}, {Tobin}  \&
  {Bourke}}{{Hull} et~al.}{2020}]{Hull+20}
{Hull} C. L.~H.,  {Le Gouellec} V. J.~M.,  {Girart} J.~M.,  {Tobin} J.~J.,
  {Bourke} T.~L.,  2020, \mn@doi [\apj] {10.3847/1538-4357/ab5809}, \href
  {https://ui.adsabs.harvard.edu/abs/2020ApJ...892..152H} {892, 152}

\bibitem[\protect\citeauthoryear{{Jones} \& {Spitzer}}{{Jones} \&
  {Spitzer}}{1967}]{JS1967}
{Jones} R.~V.,  {Spitzer} Lyman J.,  1967, \mn@doi [\apj] {10.1086/149086},
  \href {https://ui.adsabs.harvard.edu/abs/1967ApJ...147..943J} {147, 943}

\bibitem[\protect\citeauthoryear{{Kataoka}, {Machida}  \& {Tomisaka}}{{Kataoka}
  et~al.}{2012}]{Kataoka+12}
{Kataoka} A.,  {Machida} M.~N.,   {Tomisaka} K.,  2012, \mn@doi [\apj]
  {10.1088/0004-637X/761/1/40}, \href
  {https://ui.adsabs.harvard.edu/abs/2012ApJ...761...40K} {761, 40}

\bibitem[\protect\citeauthoryear{{Kataoka} et~al.,}{{Kataoka}
  et~al.}{2015}]{Kataoka15}
{Kataoka} A.,  et~al., 2015, \mn@doi [\apj] {10.1088/0004-637X/809/1/78}, \href
  {https://ui.adsabs.harvard.edu/abs/2015ApJ...809...78K} {809, 78}

\bibitem[\protect\citeauthoryear{{Kataoka}, {Muto}, {Momose}, {Tsukagoshi}  \&
  {Dullemond}}{{Kataoka} et~al.}{2016}]{Kataoka+16}
{Kataoka} A.,  {Muto} T.,  {Momose} M.,  {Tsukagoshi} T.,   {Dullemond} C.~P.,
  2016, \mn@doi [\apj] {10.3847/0004-637X/820/1/54}, \href
  {https://ui.adsabs.harvard.edu/abs/2016ApJ...820...54K} {820, 54}

\bibitem[\protect\citeauthoryear{{Ko}, {Liu}, {Lai}, {Ching}, {Rao}  \&
  {Girart}}{{Ko} et~al.}{2020}]{KoCL+20}
{Ko} C.-L.,  {Liu} H.~B.,  {Lai} S.-P.,  {Ching} T.-C.,  {Rao} R.,   {Girart}
  J.~M.,  2020, \mn@doi [\apj] {10.3847/1538-4357/ab5e79}, \href
  {https://ui.adsabs.harvard.edu/abs/2020ApJ...889..172K} {889, 172}

\bibitem[\protect\citeauthoryear{{Kuffmeier}, {Reissl}, {Wolf}, {Stephens}  \&
  {Calcutt}}{{Kuffmeier} et~al.}{2020}]{Kuffmeier+20}
{Kuffmeier} M.,  {Reissl} S.,  {Wolf} S.,  {Stephens} I.,   {Calcutt} H.,
  2020, \mn@doi [\aap] {10.1051/0004-6361/202038111}, \href
  {https://ui.adsabs.harvard.edu/abs/2020A&A...639A.137K} {639, A137}

\bibitem[\protect\citeauthoryear{{Kwon}, {Stephens}, {Tobin}, {Looney}, {Li},
  {van der Tak}  \& {Crutcher}}{{Kwon} et~al.}{2019}]{Kwon+19}
{Kwon} W.,  {Stephens} I.~W.,  {Tobin} J.~J.,  {Looney} L.~W.,  {Li} Z.-Y.,
  {van der Tak} F. F.~S.,   {Crutcher} R.~M.,  2019, \mn@doi [\apj]
  {10.3847/1538-4357/ab24c8}, \href
  {https://ui.adsabs.harvard.edu/abs/2019ApJ...879...25K} {879, 25}

\bibitem[\protect\citeauthoryear{{Lam}, {Li}, {Chen}, {Tomida}  \&
  {Zhao}}{{Lam} et~al.}{2019}]{Lam2019}
{Lam} K.~H.,  {Li} Z.-Y.,  {Chen} C.-Y.,  {Tomida} K.,   {Zhao} B.,  2019,
  \mn@doi [\mnras] {10.1093/mnras/stz2436}, \href
  {https://ui.adsabs.harvard.edu/abs/2019MNRAS.489.5326L} {489, 5326}

\bibitem[\protect\citeauthoryear{{Lazarian}}{{Lazarian}}{2007}]{Lazarian2007}
{Lazarian} A.,  2007, \mn@doi [\jqsrt] {10.1016/j.jqsrt.2007.01.038}, \href
  {https://ui.adsabs.harvard.edu/abs/2007JQSRT.106..225L} {106, 225}

\bibitem[\protect\citeauthoryear{{Le Gouellec} et~al.,}{{Le Gouellec}
  et~al.}{2019}]{LeGouellec+19}
{Le Gouellec} V. J.~M.,  et~al., 2019, \mn@doi [\apj]
  {10.3847/1538-4357/ab43c2}, \href
  {https://ui.adsabs.harvard.edu/abs/2019ApJ...885..106L} {885, 106}

\bibitem[\protect\citeauthoryear{{Le Gouellec} et~al.,}{{Le Gouellec}
  et~al.}{2020}]{LeGouellec2020}
{Le Gouellec} V.~J.~M.,  et~al., 2020, \mn@doi [\aap]
  {10.1051/0004-6361/202038404}, \href
  {https://ui.adsabs.harvard.edu/abs/2020A&A...644A..11L} {644, A11}

\bibitem[\protect\citeauthoryear{{Lee}, {Hull}  \& {Offner}}{{Lee}
  et~al.}{2017}]{Lee_Hull+17}
{Lee} J. W.~Y.,  {Hull} C. L.~H.,   {Offner} S. S.~R.,  2017, \mn@doi [\apj]
  {10.3847/1538-4357/834/2/201}, \href
  {https://ui.adsabs.harvard.edu/abs/2017ApJ...834..201L} {834, 201}

\bibitem[\protect\citeauthoryear{{Lin}, {Li}, {Yang}, {Looney}, {Lee},
  {Stephens}  \& {Lai}}{{Lin} et~al.}{2020}]{Lin+20a}
{Lin} Z.-Y.~D.,  {Li} Z.-Y.,  {Yang} H.,  {Looney} L.,  {Lee} C.-F.,
  {Stephens} I.,   {Lai} S.-P.,  2020, \mn@doi [\mnras]
  {10.1093/mnras/staa542}, \href
  {https://ui.adsabs.harvard.edu/abs/2020MNRAS.493.4868L} {493, 4868}

\bibitem[\protect\citeauthoryear{{Maury} et~al.,}{{Maury}
  et~al.}{2018}]{Maury+18}
{Maury} A.~J.,  et~al., 2018, \mn@doi [\mnras] {10.1093/mnras/sty574}, \href
  {https://ui.adsabs.harvard.edu/abs/2018MNRAS.477.2760M} {477, 2760}

\bibitem[\protect\citeauthoryear{{McKee} \& {Ostriker}}{{McKee} \&
  {Ostriker}}{2007}]{MO07}
{McKee} C.~F.,  {Ostriker} E.~C.,  2007, \mn@doi [\araa]
  {10.1146/annurev.astro.45.051806.110602}, \href
  {https://ui.adsabs.harvard.edu/abs/2007ARA&A..45..565M} {45, 565}

\bibitem[\protect\citeauthoryear{Morrish}{Morrish}{1980}]{Morrish1980}
Morrish A.~H.,  1980, The Physical Principles of Magnetism.
Wiley series on the science and technology of materials, R. E. Krieger
  Publishing Company, \url {https://books.google.com/books?id=ZjUbAQAAIAAJ}

\bibitem[\protect\citeauthoryear{{Ossenkopf} \& {Henning}}{{Ossenkopf} \&
  {Henning}}{1994}]{Ossenkopf1994}
{Ossenkopf} V.,  {Henning} T.,  1994, \aap, \href
  {https://ui.adsabs.harvard.edu/abs/1994A&A...291..943O} {291, 943}

\bibitem[\protect\citeauthoryear{{Padovani} et~al.,}{{Padovani}
  et~al.}{2012}]{Padovani+12}
{Padovani} M.,  et~al., 2012, \mn@doi [\aap] {10.1051/0004-6361/201219028},
  \href {https://ui.adsabs.harvard.edu/abs/2012A&A...543A..16P} {543, A16}

\bibitem[\protect\citeauthoryear{{Planck Collaboration Int. XIX}}{{Planck
  Collaboration Int. XIX}}{2015}]{PlanckXIX}
{Planck Collaboration Int. XIX} 2015, \mn@doi [\aap]
  {10.1051/0004-6361/201424082}, \href
  {https://ui.adsabs.harvard.edu/abs/2015A&A...576A.104P} {576, A104}

\bibitem[\protect\citeauthoryear{{Purcell}}{{Purcell}}{1979}]{P79}
{Purcell} E.~M.,  1979, \mn@doi [\apj] {10.1086/157204}, \href
  {https://ui.adsabs.harvard.edu/abs/1979ApJ...231..404P} {231, 404}

\bibitem[\protect\citeauthoryear{{Purcell} \& {Spitzer}}{{Purcell} \&
  {Spitzer}}{1971}]{PS1971}
{Purcell} E.~M.,  {Spitzer} Lyman J.,  1971, \mn@doi [\apj] {10.1086/151002},
  \href {https://ui.adsabs.harvard.edu/abs/1971ApJ...167...31P} {167, 31}

\bibitem[\protect\citeauthoryear{{Reissl}, {Wolf}  \& {Brauer}}{{Reissl}
  et~al.}{2016}]{POLARIS}
{Reissl} S.,  {Wolf} S.,   {Brauer} R.,  2016, \mn@doi [\aap]
  {10.1051/0004-6361/201424930}, \href
  {https://ui.adsabs.harvard.edu/abs/2016A&A...593A..87R} {593, A87}

\bibitem[\protect\citeauthoryear{{Roberge}, {Degraff}  \& {Flaherty}}{{Roberge}
  et~al.}{1993}]{Roberge1993}
{Roberge} W.~G.,  {Degraff} T.~A.,   {Flaherty} J.~E.,  1993, \mn@doi [\apj]
  {10.1086/173390}, \href
  {https://ui.adsabs.harvard.edu/abs/1993ApJ...418..287R} {418, 287}

\bibitem[\protect\citeauthoryear{{Sadavoy} et~al.,}{{Sadavoy}
  et~al.}{2018a}]{Sadavoy+18a}
{Sadavoy} S.~I.,  et~al., 2018a, \mn@doi [\apj] {10.3847/1538-4357/aac21a},
  \href {https://ui.adsabs.harvard.edu/abs/2018ApJ...859..165S} {859, 165}

\bibitem[\protect\citeauthoryear{{Sadavoy} et~al.,}{{Sadavoy}
  et~al.}{2018b}]{Sadavoy+18b}
{Sadavoy} S.~I.,  et~al., 2018b, \mn@doi [\apj] {10.3847/1538-4357/aaef81},
  \href {https://ui.adsabs.harvard.edu/abs/2018ApJ...869..115S} {869, 115}

\bibitem[\protect\citeauthoryear{{Sadavoy} et~al.,}{{Sadavoy}
  et~al.}{2019}]{Sadavoy19}
{Sadavoy} S.~I.,  et~al., 2019, \mn@doi [\apjs] {10.3847/1538-4365/ab4257},
  \href {https://ui.adsabs.harvard.edu/abs/2019ApJS..245....2S} {245, 2}

\bibitem[\protect\citeauthoryear{{Segura-Cox} et~al.,}{{Segura-Cox}
  et~al.}{2018}]{SeguraCox18}
{Segura-Cox} D.~M.,  et~al., 2018, \mn@doi [\apj] {10.3847/1538-4357/aaddf3},
  \href {https://ui.adsabs.harvard.edu/abs/2018ApJ...866..161S} {866, 161}

\bibitem[\protect\citeauthoryear{Shu}{Shu}{1992}]{Shu1992}
Shu F.~H.,  1992, The Physics of Astrophysics: Gas dynamics.
Series of books in astronomy, University Science Books, \url
  {https://books.google.com/books?id=50VYSc56URUC}

\bibitem[\protect\citeauthoryear{{Stephens} et~al.,}{{Stephens}
  et~al.}{2017}]{Stephens17}
{Stephens} I.~W.,  et~al., 2017, \mn@doi [\apj] {10.3847/1538-4357/aa8262},
  \href {https://ui.adsabs.harvard.edu/abs/2017ApJ...846...16S} {846, 16}

\bibitem[\protect\citeauthoryear{{Stephens} et~al.,}{{Stephens}
  et~al.}{2018}]{MASSES18}
{Stephens} I.~W.,  et~al., 2018, \mn@doi [\apjs] {10.3847/1538-4365/aacda9},
  \href {https://ui.adsabs.harvard.edu/abs/2018ApJS..237...22S} {237, 22}

\bibitem[\protect\citeauthoryear{{Stone}, {Gardiner}, {Teuben}, {Hawley}  \&
  {Simon}}{{Stone} et~al.}{2008}]{Athena2008}
{Stone} J.~M.,  {Gardiner} T.~A.,  {Teuben} P.,  {Hawley} J.~F.,   {Simon}
  J.~B.,  2008, \mn@doi [\apjs] {10.1086/588755}, \href
  {https://ui.adsabs.harvard.edu/abs/2008ApJS..178..137S} {178, 137}

\bibitem[\protect\citeauthoryear{{Takahashi}, {Machida}, {Tomisaka}, {Ho},
  {Fomalont}, {Nakanishi}  \& {Girart}}{{Takahashi}
  et~al.}{2019}]{TakahashiS+19}
{Takahashi} S.,  {Machida} M.~N.,  {Tomisaka} K.,  {Ho} P. T.~P.,  {Fomalont}
  E.~B.,  {Nakanishi} K.,   {Girart} J.~M.,  2019, \mn@doi [\apj]
  {10.3847/1538-4357/aaf6ed}, \href
  {https://ui.adsabs.harvard.edu/abs/2019ApJ...872...70T} {872, 70}

\bibitem[\protect\citeauthoryear{{Tazaki}, {Lazarian}  \& {Nomura}}{{Tazaki}
  et~al.}{2017}]{Tazaki+17}
{Tazaki} R.,  {Lazarian} A.,   {Nomura} H.,  2017, \mn@doi [\apj]
  {10.3847/1538-4357/839/1/56}, \href
  {https://ui.adsabs.harvard.edu/abs/2017ApJ...839...56T} {839, 56}

\bibitem[\protect\citeauthoryear{{Tobin} et~al.,}{{Tobin}
  et~al.}{2015}]{2015ApJ...805..125T}
{Tobin} J.~J.,  et~al., 2015, \mn@doi [\apj] {10.1088/0004-637X/805/2/125},
  \href {https://ui.adsabs.harvard.edu/abs/2015ApJ...805..125T} {805, 125}

\bibitem[\protect\citeauthoryear{{Tobin} et~al.,}{{Tobin}
  et~al.}{2016}]{Tobin16}
{Tobin} J.~J.,  et~al., 2016, \mn@doi [\apj] {10.3847/0004-637X/818/1/73},
  \href {https://ui.adsabs.harvard.edu/abs/2016ApJ...818...73T} {818, 73}

\bibitem[\protect\citeauthoryear{{Valdivia}, {Maury}, {Brauer}, {Hennebelle},
  {Galametz}, {Guillet}  \& {Reissl}}{{Valdivia} et~al.}{2019}]{Valdivia+19}
{Valdivia} V.,  {Maury} A.,  {Brauer} R.,  {Hennebelle} P.,  {Galametz} M.,
  {Guillet} V.,   {Reissl} S.,  2019, \mn@doi [\mnras] {10.1093/mnras/stz2056},
  \href {https://ui.adsabs.harvard.edu/abs/2019MNRAS.488.4897V} {488, 4897}

\bibitem[\protect\citeauthoryear{{Yang}}{{Yang}}{2021}]{Yang21}
{Yang} H.,  2021, arXiv e-prints, \href
  {https://ui.adsabs.harvard.edu/abs/2021arXiv210310243Y} {p. arXiv:2103.10243}

\bibitem[\protect\citeauthoryear{{Yang}, {Li}, {Looney}  \& {Stephens}}{{Yang}
  et~al.}{2016}]{Yang16}
{Yang} H.,  {Li} Z.-Y.,  {Looney} L.,   {Stephens} I.,  2016, \mn@doi [\mnras]
  {10.1093/mnras/stv2633}, \href
  {https://ui.adsabs.harvard.edu/abs/2016MNRAS.456.2794Y} {456, 2794}

\bibitem[\protect\citeauthoryear{{Yang}, {Li}, {Looney}, {Girart}  \&
  {Stephens}}{{Yang} et~al.}{2017}]{Yang+17}
{Yang} H.,  {Li} Z.-Y.,  {Looney} L.~W.,  {Girart} J.~M.,   {Stephens} I.~W.,
  2017, \mn@doi [\mnras] {10.1093/mnras/stx1951}, \href
  {https://ui.adsabs.harvard.edu/abs/2017MNRAS.472..373Y} {472, 373}

\bibitem[\protect\citeauthoryear{{Yen} et~al.,}{{Yen} et~al.}{2020}]{Yen+20}
{Yen} H.-W.,  et~al., 2020, \mn@doi [\apj] {10.3847/1538-4357/ab7eb3}, \href
  {https://ui.adsabs.harvard.edu/abs/2020ApJ...893...54Y} {893, 54}

\bibitem[\protect\citeauthoryear{{Zucker}, {Speagle}, {Schlafly}, {Green},
  {Finkbeiner}, {Goodman}  \& {Alves}}{{Zucker} et~al.}{2019}]{Zucker19}
{Zucker} C.,  {Speagle} J.~S.,  {Schlafly} E.~F.,  {Green} G.~M.,  {Finkbeiner}
  D.~P.,  {Goodman} A.~A.,   {Alves} J.,  2019, \mn@doi [\apj]
  {10.3847/1538-4357/ab2388}, \href
  {https://ui.adsabs.harvard.edu/abs/2019ApJ...879..125Z} {879, 125}

\makeatother
\end{thebibliography}




 \appendix
 \section{Deprojected Polarization Results for Individual Sources}
\label{app:reproj}

In this appendix, we describe the procedure for deprojecting the dust continuum map to obtain the distributions of the polarization fraction and orientation as a function of the radius for
8 of the 10 protostellar systems reported in \citet{Cox18}. As discussed in Section~\ref{sec:ALMAdata}, we define a disk plane for each source based on its outflow direction inferred from the MASSES survey \citep{Stephens17,MASSES18}). The inclination angle of a system to the plane of the sky could be more uncertain unless a well-defined dust disk is observed and fitted. This is the case for 3 of our 8 targets using the analysis from the VLA Nascent Disk and Multiplicity Survey of Perseus Protostars (VANDAM; \citealt{SeguraCox18}). We thus use the protostellar disks-fitted inclination angles for these three systems (Per-emb-11, Per-emb-14, Per-emb-50). For the remaining protostellar systems without well-established disks, we adopt a single value of $45^\circ$; the exact value adopted for the inclination angle does not change our conclusions qualitatively.

To compare the polarization data from disks with different sizes, we normalize the distance from the center by a characteristic radius $R_{\rm 10\%}$.
As described in Section~\ref{sec:ALMAdata}, it is the radius of a circle that encloses the same area as the ellipse at 10\% maximum intensity from 2D Gaussian fitting of the ALMA Band 7 Stokes I image.
It is not our intention to claim that $R_{\rm 10\%}$ represents the actual disk radius, although the transition from disk-scale to envelope-scale polarization patterns does seem to happen between $\sim 1-2~R_{\rm 10\%}$ for all protostellar systems investigated in this study (see e.g.,~Fig.~\ref{fig:ObsTrends}).

\begin{figure*}
    \centering
    \includegraphics[width=0.9\textwidth]{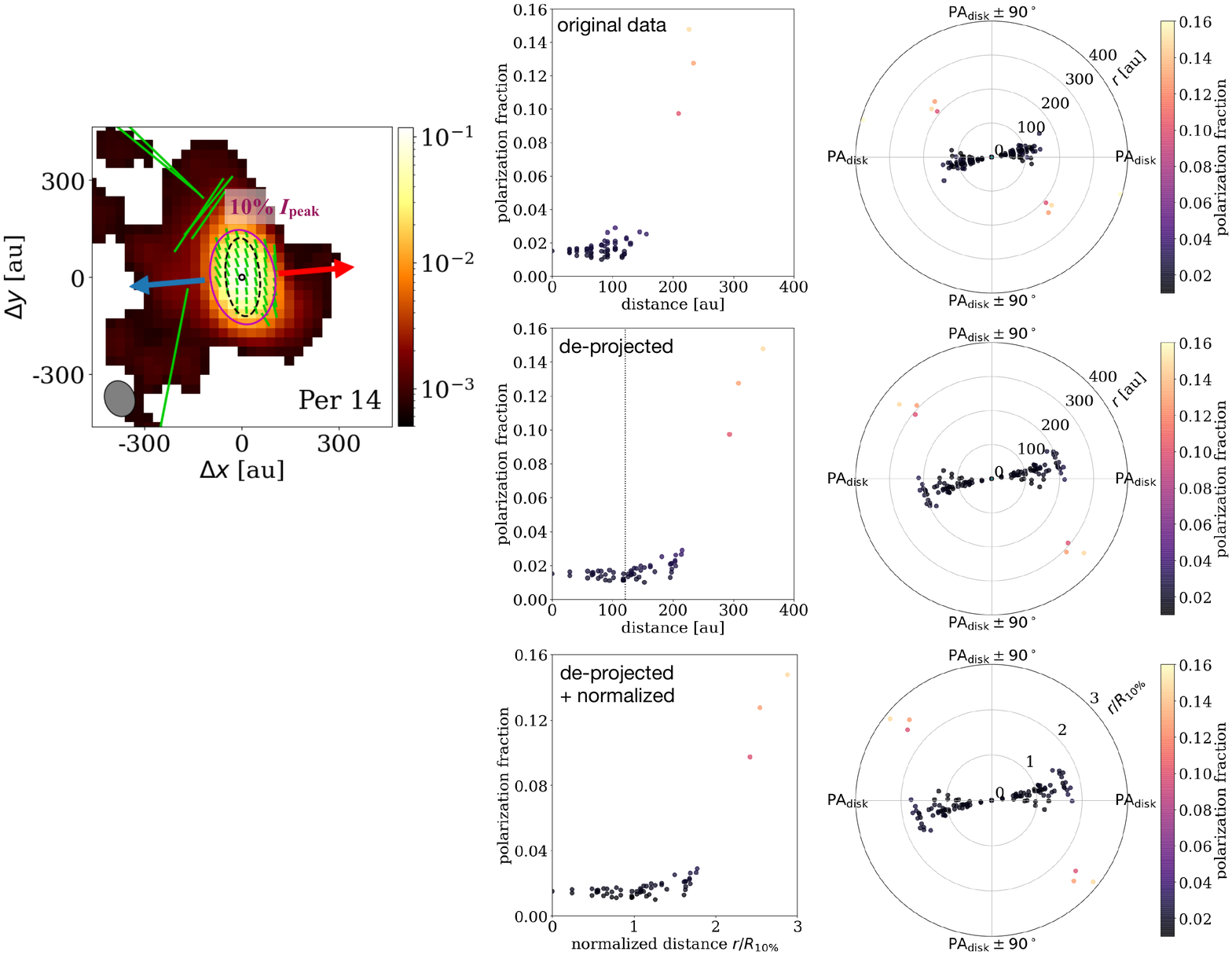}
    \caption{Demonstration of the deprojection and renormalization effects on observation data, using Per\,14 as the example. {\it Left:} total intensity map (in Jy/beam) and polarization (B-)vectors from ALMA Band 7 observation presented in \citet{Cox18}, with outflow directions indicated by blue and red arrows and beam in the lower left corner. The Gaussian-fitted $10\%$ peak intensity contour ({\it purple ellipse}) and the inferred disk adopted in the deprojection process (radius $= R_{\rm 10\%}$ and inclination $=45^\circ$; {\it black dashed ellipse}) are also marked. {\it Middle and right:} similar to Fig.~\ref{fig:ObsTrends} but for the original ({\it top}), deprojected ({\it middle}), and deprojected and normalized ({\it bottom}) data of Per\,14. The polarization fraction is color coded, with the color bar shown on the right.}
    \label{fig:deprojdemo}
\end{figure*}

The effects of the deprojection and radius normalization are illustrated in Fig.~\ref{fig:deprojdemo} for Per-emb-14. The middle column shows the polarization fraction as a function of distance to the center of the system (defined as the location of the peak intensity), both before (top) and after (middle) the deprojection, and after the normalization with $R_{\rm 10\%}$ (bottom).
The normalization suggests that the transition from flat (disk-scale) to steep (envelope-scale) $p-r$ correlation happens roughly at $R_{\rm 10\%}$ (vertical dotted line in the middle panel).
The right column of Fig.~\ref{fig:deprojdemo} shows the corresponding polar plot of the polarization orientations.
Note that our deprojection routine only provides a correction on the distance to the system center and does not modify the orientation of the polarization segments; this is the reason why the distributions of polarization orientations in the polar plots (the right panels of Fig.~\ref{fig:deprojdemo}) are barely impacted by the deprojection process.

 \begin{table}
  \caption{Deprojection parameters for individual sources}
  \label{tab:reproj}
  \begin{tabular}{lcclll}
    \hline
    \hline
    Target & PA\tablefootmark{a} & inclination\tablefootmark{b} & $R_{\rm 10\%}$ & regridded & regridded \\
    & $(^\circ)$ & $(^\circ)$ & (au) & $0.1^{\prime\prime}$ & $0.2^{\prime\prime}$\\
    \hline
    Per\,2  & 39 &  45 & 311 & 301 & 313 \\
    Per\,5 & 35 & 45  & 126 & 126 & 139 \\
    Per\,11  & 72 & 44\tablefootmark{c} & 138 & 137 & 154 \\
    Per\,14 & 5 & 64\tablefootmark{c}  & 123 & 121 & 134\\
    Per\,18  & 60 & 45 & 161 & 158 & 166 \\
    Per\,26  & 72 & 45 & 109 & 109 & 125\\
    Per\,29 & 42 & 45  & 117 & 117 & 136 \\
    Per\,50  & 14 & 67\tablefootmark{c} & 103 & 103 & 126 \\
    \hline
  \end{tabular}
  \tablefoot{
    \tablefoottext{a}{The disk major axis is taken to be perpendicular to the outflow direction given in \citet{MASSES18}. }
    \tablefoottext{b}{Assumed to be $45^\circ$ unless otherwise noted.}
    \tablefoottext{c}{From \citet{SeguraCox18}.}
  }
\end{table}

\begin{figure*}
    \centering
    \includegraphics[width=0.72\textwidth]{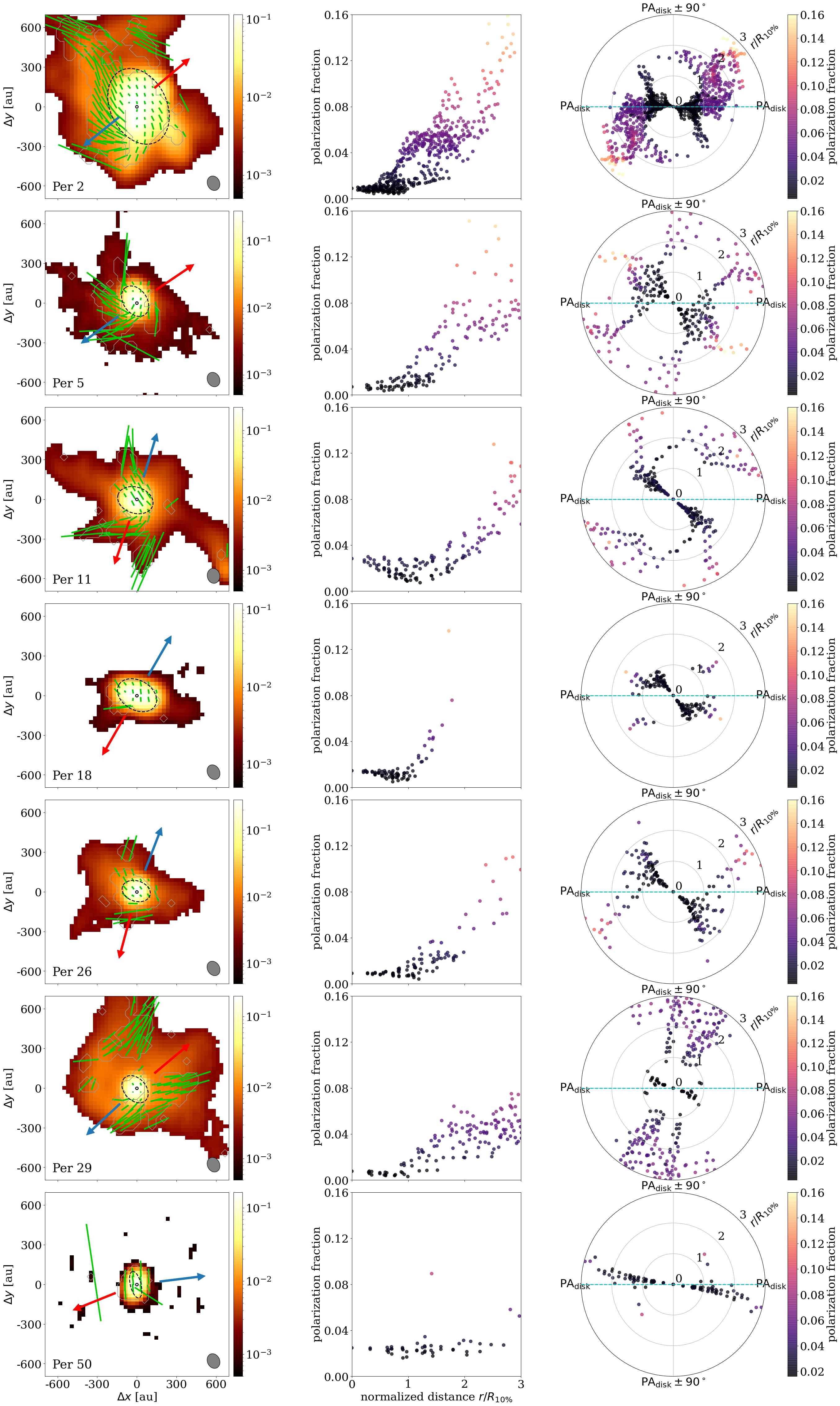}
    \caption{Summary of the ALMA Band 7 polarization observations toward the 8 protostellar systems investigated in this study (with the exception of Per\,14) and the deprojection results. Plotted are the ALMA Band 7 polarization B-vector map ({\it left column}) superposed on the dust continuum (background color map in Jy/beam with mask $I > 5\sigma_I$, outflow direction indicated by red and blue arrows and beam in the lower right corner), the scatter plots of the polarization fraction as a function of the normalized distance ({\it middle column}; darker color means more pixels), and the scatter plots of the B-vector orientations ($0^\circ$ for north, {\it right column}) with the ``disk plane'' defined from the outflow direction marked as the dashed horizontal cyan line (i.e., the polar plots are rotated so that the position angle of the disk is horizontal). The polarization fraction is color coded, with the color bar shown on the right.}
    \label{fig:obsprhist}
\end{figure*}

Fig.~\ref{fig:obsprhist} summarizes the deprojection results for the 8 ALMA targets considered in this study except for Per-emb-14, which is presented in Fig.~\ref{fig:deprojdemo}. The deprojection parameters (position angle, inclination angle, normalized radius $R_{\rm 10\%}$) for individual systems are listed in Table~\ref{tab:reproj}.
The disk-to envelope transition of the $p-r$ correlation at the distance $\sim 1-2 R_{\rm 10\%}$ is clear in all systems.
We would like to point out that such transition is not an artificial result of our deprojection routine; in fact, it could also be seen even before deprojection (see e.g.,~Fig.~\ref{fig:deprojdemo}), and the deprojection simply highlights this transition.

\section{Extinction of Polarized Emission}
\label{app:extinction}

To properly account for the extinction of polarized emission, one needs to solve the vector radiation transfer equation:
\begin{equation}
    \frac{d}{\rho\,ds} \mathbf{S} = -\mathcal{K}\,\mathbf{S} + B_\nu(T)\,\mathbf{a}.
    \label{eq:vrt}
\end{equation}
where $\mathbf{S} = (I,Q,U)$ is the radiation (Stokes) vector, $s$ the distance along the light path, and $B_\nu(T)$ the Planck function. The extinction matrix $\mathcal{K}$ and absorption vector $\mathbf{a}$ are related to the extinction and polarization opacities, $C_e$ and $C_p$, by \citep[see e.g. the POLARIS code][]{POLARIS}
\begin{subequations}
\begin{align}
    & \mathcal{K} =
    \left(~\begin{matrix}
        C_e            & C_p \cos 2\psi & C_p \sin 2\psi \\
        C_p \cos 2\psi & C_e            & 0              \\
        C_p \sin 2\psi & 0              & C_e
    \end{matrix}~\right), \\
    & \mathbf{a} =
    \left(~\begin{matrix}
        C_e            & C_p \cos 2\psi & C_p \sin 2\psi
    \end{matrix}~\right)^T,
\end{align}
\end{subequations}
where $\psi$ is the angle between the magnetic field and the direction of positive $Q$ in the sky plane. Taking into account of the magnetic alignment condition, as specified by the magnetic alignment probability $A$, defined in equation~(\ref{eqn:frac}), we have the extinction and polarization opacities given by:
\begin{subequations}
\begin{align}
    C_e &= \kappa\left\{ 1 - \alpha \left[ A \left( \frac{\cos^2\gamma}{2}-\frac{1}{3} \right) + \frac{1-A}{6} \right] \right\}, \\
    C_p &= \kappa\,\alpha\,A\,\cos^2\gamma,
\end{align}
\end{subequations}
where $\kappa$ is the dust opacity (cross-section per gram of gas rather than dust), which is taken to be $\kappa=1.75\times 10^{-2}\,\mathrm{cm^2/g}$ (assuming a gas-to-dust mass ratio of 100) at Band 7 \citep[][logarithmically interpolated at 870 $\mu$m]{Ossenkopf1994}.

One way to solve the vector radiative transfer equation~(\ref{eq:vrt}) is through the formal solution for the Stokes vector
\begin{equation}
    \mathbf{S} = \int \mathcal{T}(s)\,\frac{\mathbf{a} + \mathbf{a}_\mathrm{sca}}{\kappa}\,\rho\,ds
    \label{eq:FormalSolution}
\end{equation}
 where the vector
 \begin{equation}
     \mathbf{a}_\mathrm{sca} = \left(~\begin{matrix}
      0 & -p_\mathrm{sca}\,(1-A)\,C_e & 0
      \end{matrix}~\right)^T
      \label{eq:scattering}
 \end{equation}
 accounts for the scattering-induced polarization included in one of the models in those simulation cells where the magnetic alignment condition is not met (i.e., $A=0$), and the matrix $\mathcal{T}(s)$ is obtained from  the integral
 \begin{equation}
     \mathcal{T}(s) = \Pi_s^\infty e^{-\mathcal{K}(s')ds'}
 \end{equation}
where $\Pi_s^\infty$ denotes the (order-preserved) geometric integration along the light path from the point of interest ($s$) to the ``observer'' at infinite distance of a function with a matrix exponential. Note that the Stokes parameters (I, Q, U) obtained from equation~(\ref{eq:FormalSolution}) at a given frequency $\nu$ are normalized by the product $\kappa B_\nu(T)$ at the frequency, as in equation~(\ref{eq:synpolideal}) for the optical thin limit.
Indeed, equation~(\ref{eq:FormalSolution}) reduces back to equation~(\ref{eq:synpolideal}) by setting $A=1$ (magnetically aligned everywhere) and $\kappa \rightarrow 0$ (optically thin).

\bsp	
\label{lastpage}
\end{document}